\documentclass[a4paper,11pt]{article}
\pdfoutput=1 

\usepackage{jheppub} 
\usepackage{amsmath,amssymb,amsthm,cancel}
\usepackage{graphicx}
\usepackage{subcaption}
\usepackage[percent]{overpic}
\usepackage[dvipsname]{xcolor}
\usepackage{multirow}
\usepackage{bbm}
\usepackage{todonotes}
\usepackage{enumitem}
\usepackage{hyperref}

\usetikzlibrary{arrows,positioning,cd,calc,math,decorations.pathreplacing,decorations.markings,decorations.pathmorphing
}
\tikzset{wave/.style={decorate, decoration=snake}}
\def\nudge{.5}
\tikzset{axis/.style={ultra thick, Red!75!black, -latex, shorten <=-\nudge cm, shorten >=-2*\nudge cm}}
\tikzset{line/.style={ultra thick,black}}

\usepackage{xcolor}
\definecolor{MyBlue}{rgb}{0.25,0.5,0.75}
\colorlet{NextBlue}{MyBlue!20}
\colorlet{SecondBlue}{MyBlue!40}

\def\be{\begin{equation}}
\def\ee{\end{equation}}

\newcommand{\CS}{\mathcal{S}}

\newcommand{\CM}{\mathcal{M}}
\newcommand{\CN}{\mathcal{N}}
\newcommand{\CC}{\mathcal{C}}
\newcommand{\CQ}{\mathcal{Q}}
\newcommand{\CT}{\mathcal{T}}

\newcommand{\IC}{\mathbb{C}}
\newcommand{\IP}{\mathbb{P}}
\newcommand{\IZ}{\mathbb{Z}}
\newcommand{\IF}{\mathbb{F}}
\newcommand{\IR}{\mathbb{R}}
\newcommand{\IU}{\mathbb{U}}

\newcommand{\ch}{\mathrm{ch}}
\newcommand{\Td}{\mathrm{Td}}
\newcommand{\Cr}{\mathrm{Cr}}
\newcommand{\rk}{\mathrm{rk}\,}

\renewcommand{\(}{\left(}
\renewcommand{\)}{\right)}
\newcommand{\Pic}{\text{Pic}}
\newcommand{\Aut}{\mathrm{Aut}}
\newcommand{\Out}{\mathrm{Out}}

\newtheorem{remark}{Remark}

\title{Quiver symmetries and  wall-crossing invariance}

\author[a,b]{Fabrizio  Del Monte}
\author[c]{Pietro Longhi}

\affiliation[a]{Centre de Recherches Math\'ematiques, Universit\'e de Montr\'eal, C. P. 6128, Succ. Centre Ville, Montr\'eal, QC H3C 3J7 Canada}

\affiliation[b]{Department of Mathematics and Statistics, Concordia University, 1455 de Maisonneuve Blvd. W. Montr\'eal, QC H3G 1M8 Canada}

\affiliation[c]{Institute for Theoretical Physics, ETH Zurich, 8093, Zurich, Switzerland}

\emailAdd{
delmonte@crm.umontreal.ca,
longhip@phys.ethz.ch, 
}

\abstract{ 
We study the BPS particle spectrum of five-dimensional superconformal field theories (SCFTs) on $\mathbb{R}^4\times S^1$ with one-dimensional Coulomb branch, by means of their associated BPS quivers. By viewing these theories as arising from the geometric engineering within M-theory, the quivers are naturally associated to the corresponding local Calabi-Yau threefold.  We show that the symmetries of the quiver, descending from the symmetries of the Calabi-Yau geometry, together with the affine root lattice structure of the flavor charges, provide equations for the Kontsevich-Soibelman wall-crossing invariant. We solve these equations iteratively: the pattern arising from the solution is naturally extended to an exact conjectural expression, that we provide for the local Hirzebruch $\mathbb{F}_0$, and local del Pezzo $dP_3$ and $dP_5$ geometries. Remarkably, the BPS spectrum consists of two copies of suitable $4d$ $\mathcal{N}=2$ spectra, augmented by Kaluza-Klein towers.
}

\begin{document} 

\maketitle
\flushbottom

\section{Introduction}

The BPS spectral problem in supersymmetric Quantum Field Theories with eight supercharges is a rich subject, with connections to various branches of theoretical and mathematical physics, such as the WKB approximation and  Stokes phenomena for Schr\"odinger equations (and higher order analogues), 
Donaldson-Thomas invariants in enumerative geometry, 
and integrable systems  both of continuous and discrete types. 
The past decade has brought remarkable progress on the case of four-dimensional $\mathcal{N}=2$ theories \cite{Gaiotto:2008cd, Gaiotto:2009hg, Gaiotto:2012rg, Alim:2011ae,  Alim:2011kw, Manschot:2010qz, Manschot:2011xc, Manschot:2012rx, Manschot:2013sya, Manschot:2014fua}, driven by seminal advances on wall-crossing phenomena \cite{Kontsevich:2008fj, Joyce:2008pc}.
More recently, various groups have started to tackle the problem of describing the spectrum of BPS states of five-dimensional SCFTs, arising from the compactification of M-theory on local Calabi-Yau threefolds, where the BPS states are to be understood as branes wrapping compact cycles of the Calabi-Yau \cite{Eager:2016yxd, Banerjee:2018syt, Banerjee:2019apt, Banerjee:2020moh, Closset:2019juk, Bousseau:2019ift,  Alexandrov:2018iao, Arguz:2021zpx, Beaujard:2020sgs, Mozgovoy:2020has, Mozgovoy:2021iwz, Descombes:2021snc}. These are the theories we will discuss in the present paper.

Both in four and five dimensions, a key feature of BPS spectra is the wall-crossing  phenomenon. 
As we move along the moduli space of vacua of a theory, the BPS spectrum is only piecewise constant, with discontinuities on codimension-one walls. 
These walls are loci where stable particles can decay, and they subdivide the moduli space into chambers with different BPS spectra in each of them.  
Even though the BPS spectrum itself is not preserved on the whole moduli space, there are still wall-crossing invariant quantities, that are constant and chamber-independent. An important example is the Konstevich-Soibelman (KS) invariant $\IU$, also known as quantum monodromy or motivic spectrum generator \cite{Kontsevich:2008fj}. 
On the one hand, it is always possible to derive the BPS spectrum in any chamber from $\IU$. 
On  the other hand, $\IU$ can be computed from knowledge of the BPS spectrum at some point in moduli space.
Thanks to its invariance property, it is sufficient to compute $\IU$ in a specific chamber where the spectrum admits a simple description, in order to obtain the spectrum in any other chamber. 
Our goal will be to compute the spectrum generator for certain five-dimensional SCFTs, by using symmetries of the underlying Calabi-Yau geometry.\footnote{For BPS spectra of 4d $\CN=2$ theories of class $\CS$, this task has been accomplished in \cite{Gaiotto:2009hg} for theories of type  $A_1$  and a general approach for all class $\CS$ theories was developed in  \cite{Longhi:2016wtv}. The latter approach should generalize to 5d $\CN=1$ theories via exponential networks, as observed in \cite{Banerjee:2019apt, Banerjee:2020moh}. However, this requires finding a  special point in moduli space (the Roman  locus of \cite{Gabella:2017hpz}). Existence of such a locus has been discussed recently in the literature \cite{Closset:2021lhd}, it would be interesting to pursue this approach for more general models than those considered here.}

Our main tool will be the so-called BPS quiver of the theory \cite{Douglas:1996sw, Douglas:2000ah, Douglas:2000qw, Denef:2002ru, Alim:2011ae,Alim:2011kw, Chuang:2013wt}. This can be thought of as the quiver characterizing the Supersymmetric Quantum Mechanics describing the low-energy dynamics of the BPS states of the theory. The spectrum is then  encoded by the associated representation theory for suitable stability  conditions.
As it turns out, this approach is  generically difficult for quivers of the five-dimensional theories considered in this paper, see e.g.  \cite{Closset:2019juk} for a recent discussion. 

An alternative way to  obtain the BPS spectrum from a quiver, which has been  used   to great  effect in the four-dimensional case, is the so-called ``mutation method" \cite{Alim:2011ae,Alim:2011kw}. 
The idea behind it is that every node in the BPS quiver represents a hypermultiplet BPS state, characterized by a ray in the complex plane of central charges. 
Since the central charge of an antiparticle is the opposite of the central charge of the corresponding particle, the whole spectrum is encoded by the rays within a half-plane. A choice of half-plane corresponds  to a choice of quiver description of the BPS spectrum, for a given point  in the  moduli space of stability conditions.
If one starts to tilt the choice of half-plane, at some point the ray of a BPS state will exit the half-plane, and  this    will  induce a change in the quiver description corresponding to a mutation of the BPS quiver at the corresponding node \cite{Berenstein:2002fi}. 
We have however done nothing physically meaningful, so the new quiver just corresponds to a dual description of the same physics, so that the charges of the new nodes of the quiver must have been also in the original spectrum. By iterating this procedure, one produces stable hypermultiplet states in a given chamber: multiplets with higher spin content appear as limiting vectors of infinite sequences of mutations, and different chambers in moduli space will correspond to different relative orderings of the phases of central charges.
In chambers  where all  stable BPS  states correspond  to hypermultiplets, the  mutation  method exhausts the whole BPS spectrum. This is sometimes  the case for BPS quivers of four-dimensional $\CN=2$ theories, however it is never the case for five-dimensional SCFTs.

While the mutation method is not as powerful in five dimensions as it is in  four dimensions, it is nevertheless quite useful in favourable circumstances.
On the one hand, it was observed in \cite{Closset:2019juk} that certain SCFTs admit choices of stability conditions for which the mutation method \emph{almost} captures the full BPS spectrum, leaving out only BPS states with  central  charges aligned along a single ray in the complex plane.
Moreover, it was proposed in \cite{Bonelli:2020dcp} that the mutation method should be generalized by exploiting symmetries of the quivers, that can be used to define discrete integrable systems giving rise to q-Painlev\'e equations \cite{Goncharov:2011hp,Fock:2014ifa,Bershtein:2017swf,mizuno2020q}, whose connection with five-dimensional theories was first recognized in a different but related context in \cite{Bershtein:2016aef,Bonelli:2017gdk}. 
In \cite[Appendix]{Longhi:2021qvz} the symmetries of the quiver were used to complete the description of BPS states derived  via mutations in \cite{Closset:2019juk}, for the case of local  $\IF_0$.
In this paper we elaborate on these ideas in  a  more precise and systematic way, and apply them to compute the motivic spectrum generator for certain local Calabi-Yau threefolds whose BPS charges have an underlying structure of  an affine root lattice, as it happens for all local del Pezzos\footnote{Local del Pezzos are open Calabi-Yau threefolds arising as total space of the canonical bundle over a del Pezzo surface. The del Pezzo surface $dP_n$ is a two-dimensional complex surface obtained by blowing up $\mathbb{P}^2$ at $n$ generic points.}. We will focus on  the specific examples local $\mathbb{F}_0$, and the local del Pezzos $dP_3,dP_5$, the latter example being non-toric.

Our main result will be to show that symmetries of a local CY geometry (not necessarily toric) together with wall-crossing invariance, are sometimes enough to constrain the BPS spectrum entirely, leading to exact expressions for $\IU$ and the spectrum, as well as highly nontrivial wall-crossing identities. 
Two important caveats apply. First, wall-crossing  invariance only constrains BPS states whose charges are not pure-flavor, i.e. whose Dirac-Schwinger-Zwanziger (DSZ) pairing with other charges is not identically zero. Our method does not determine  BPS states with pure-flavor charges (such as boundstates of D0 branes), these must be obtained by other methods. Our results should be regarded as partially conjectural, admitting  the possibility of additional BPS states with pure-flavor charges that we have not detected.\footnote{For the  case of local $\IF_0$ we believe the  result to be exhaustive, based on  comparisons with exponential networks in  \cite{Longhi:2021qvz}. More generally, the  ambiguity on pure-flavor states may be related to  the fact that our approach is independent of  a  choice of superpotential associated to a quiver. 
It is expected that different choices of superpotentials would result in different BPS spectra. 
The difference should then lie in the spectrum of pure-flavor states, while the part of the  spectrum that we compute should be independent of such a choice.} 
Second, we use symmetries of the quiver and wall-crossing invariance to derive exact equations for $\IU$, which we solve iteratively, leading to partial descriptions of the BPS spectrum. We  then  complete these descriptions by observing certain patterns in the spectrum, namely certain periodicities, leading to a \emph{conjectural} expression for $\IU$. 

With these caveats in  mind, we arrive at explicit and exact (conjectural) descriptions of the BPS spectrum. 
The starting point for our approach is the observation that the so-called Cremona group $\Cr(dP_n)$ of del Pezzo surfaces $dP_n$ \cite{2001CMaPh.220..165S}, are realized by the extended affine Weyl group $\widetilde{W}(E_n^{(1)})$ (there can be an extra finite group factor, see Section \ref{Sec:StabRoots} for details) acting on the BPS charge lattice of the theory through sequences of mutations and permutations that leave the quiver invariant. 
These transformations form the automorphism group of the quiver $\Aut_Q$, so that
\begin{equation}
\Aut_Q\approx\Cr(dP_n)\supseteq\widetilde{W}(E_n^{(1)})=\left(W(E_n)\ltimes \CT(E_n^{(1)})\right)\times\Out(E_n^{(1)}),
\end{equation}
where we decomposed the extended affine Weyl group into affine translations $\CT$, simple reflections $W$, and Dynkin diagram automorphisms $\Out(E_n^{(1)})$. Among these transformations, a special role is taken by affine translations $\CT(E_n^{(1)})$, that can be thought of as discrete time evolutions on the BPS charges, and by the subgroup $\Pi_Q$ of quiver automorphisms consisting only of permutations on the nodes.\footnote{In the context of 4d $\CN=2$ theories, the time evolution of discrete integrable systems associated to BPS quivers was related to the study of BPS spectra in \cite{Cecotti:2014zga,Cirafici:2017iju,Cirafici:2020qlf}}

We will show that it is possible to identify certain affine translations with tiltings of the upper half-plane corresponding to an appropriate stability condition, so that we associate to an affine translation a chamber in the moduli space of the theory. Furthermore, we will show that the chambers obtained these way have the ``collimation''property: this means that their spectrum is organized into towers of hypermultiplets, with all the higher-spin BPS states aligned on the real axis of the complex plane of BPS central charges. The  motivic spectrum generator in such a chamber is factorized as
\begin{equation}\label{eq:UFactor}
\IU = \IU(\measuredangle^+)\cdot  \IU(\IR^+)\cdot  \IU(\measuredangle^-),
\end{equation}
where $\IU(\measuredangle^{\pm})$ are the contributions from BPS states with central charges lying inside the first quadrant and fourth quadrant respectively, while $\IU(\IR^+)$ contains the contribution of the rays lying on the positive real axis. We will show that the affine translations fully determine $\IU(\measuredangle^{\pm})$, while $\IU(\IR^+)$ can be fixed using the remaining transformations of $Aut_Q\approx\Cr(dP_n)$. This is done by noting that the permutation symmetry group $\Pi_Q\subset\Aut_Q$ of the quiver contains transformations that are automorphisms of the Dynkin diagram. Denoting such a transformation by $\pi$, this has the effect of relating different affine translations $T,T'$ as
\begin{equation}
T'=\pi^{-1}T\pi.
\end{equation}
As a result, the collimation chambers $\CC$, $\CC'$ constructed from the two translations $T$ and $T'$ will be simply related by the permutation $\pi$ of the charge vectors. This statement, combined with wall-crossing invariance and the factorization \eqref{eq:UFactor}, allows us to compute $\IU(\IR^+)$ recursively from
\begin{equation}\label{eq:WC-identity-intro}
\IU[\CC]=\IU[\CC']=\IU[\pi(\CC)]\,.
\end{equation}
For the cases we consider, the affine translations naturally split the BPS in quiver into two four-dimensional quivers: this was first noticed in \cite{Bonelli:2020dcp}, and is shown in Figure \ref{fig:4d-subquivers}. As we will discuss shortly, this is reflected in the full BPS spectrum.

\subsection*{Main  results}
Solving equation (\ref{eq:WC-identity-intro}) for suitable chambers of local $\IF_0, dP_3, dP_5$  we obtain the following conjectural BPS spectra.\footnote{
As mentioned earlier, BPS states with pure-flavor charges go undetected by  the  wall-crossing identity (\ref{eq:WC-identity-intro}). These states, which include boundstates of D0 branes, can't be determined by wall-crossing invariance, since such states do not participate in wall-crossing.
}

\begin{figure}[h]
\begin{center}
\begin{subfigure}{.35\textwidth}
\centering
\includegraphics[width=\textwidth]{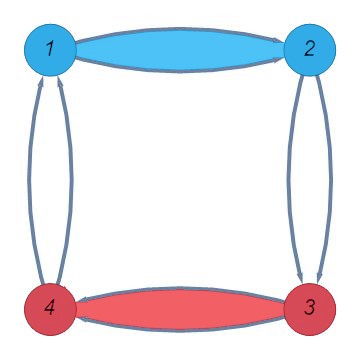}
\caption{Four-dimensional subquivers for the flow $T_1$ of $\mathbb{F}_0$. The towers \eqref{eq:T1-F0-charges} are two copies of the dyons for 4d $SU(2)$ pure SYM, with additional $D0$ towers over the 4d vector multiplets.}
\label{Fig:F0-intro}
\end{subfigure}\hspace*{.15\textwidth}
\begin{subfigure}{.35\textwidth}
\centering
\includegraphics[width=\textwidth]{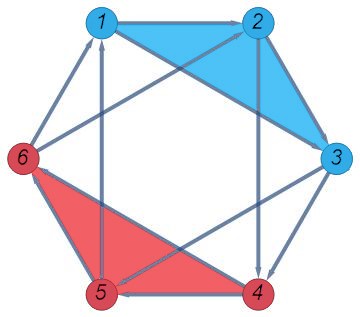}
\caption{Four-dimensional subquivers for the flow $T_1$ of $dP_3$. The towers \eqref{eq:T1T2-dP3-charges} are two copies of the dyons for 4d $SU(2)$ SYM with $N_f=1$, with the addition of $D0$ towers over the 4d quarks and vector multiplets.}
\label{Fig:dP3-intro}
\end{subfigure}
\begin{subfigure}{.35\textwidth}
\centering
\includegraphics[width=\textwidth]{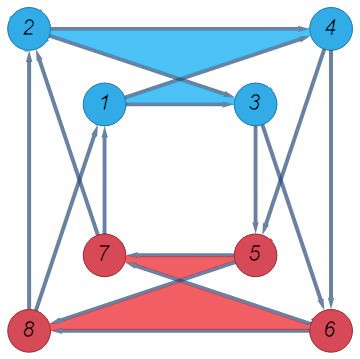}
\caption{Four-dimensional subquivers for the flow $T_5$ of $dP_5$. The towers \eqref{eq:dP5Transl1} are two copies of the dyons for 4d $SU(2)$ SYM with $N_f=2$, with the addition of $D0$ towers over the 4d quarks and vector multiplets.}
\label{Fig:dP5-intro}
\end{subfigure}
\caption{}
\label{fig:4d-subquivers}
\end{center}
\end{figure}

For local $\IF_0$ we choose generators  $\gamma_i$ of the charge lattice corresponding to nodes of the BPS quivers shown in  Figure~\ref{Fig:F0-intro}.
In a chamber where
\be
	Z_{\gamma_1} = Z_{\gamma_3} \,,
	\quad
	Z_{\gamma_2} = Z_{\gamma_4} \,,
	\quad
	\arg Z_{\gamma_1} > \arg Z_{\gamma_2}\,,
	\quad
	Z_{\gamma_1}+Z_{\gamma_2}\in \IR^+\,,
\ee
we find the following BPS spectrum 
\begin{equation}
\begin{array}{|c|c|}
	\hline
	\gamma & \Omega(\gamma;y) \\
	\hline\hline
	\gamma_1 + k (\gamma_1+\gamma_2) & 1\\
	-\gamma_1 + (k+1) (\gamma_1+\gamma_2) & 1\\
	\gamma_3+ k (\gamma_3+\gamma_4) & 1\\
	-\gamma_3+ (k+1) (\gamma_3+\gamma_4) & 1\\
	\hline
	\gamma_1+\gamma_2+k\gamma_{D0} & y+y^{-1} \\
	-\gamma_1-\gamma_2+(k+1)\gamma_{D0} & y+y^{-1} \\
	(k+1)\gamma_{D0} & y^3 + 2y+y^{-1}\\
	\hline
\end{array}
\end{equation}
where $\gamma_{D0} = \sum_{i=1}^{4} \gamma_i$. 
The spectrum  also  includes the respective antiparticles obtained by sending $\gamma\to-\gamma$. Note that this spectrum consists of two copies of the weakly coupled spectrum of pure $SU(2)$ SYM in 4d \cite{Alim:2011kw}, with the extra KK towers in lower part of the table.
We also observe that the spectrum  is  organized into towers over $\gamma_1,\gamma_3$ and $\gamma_1+\gamma_2, \gamma_{D0}$ with steps $\gamma_1+\gamma_2, \gamma_3+\gamma_4$ and $\gamma_{D0}$. This structure is strongly reminiscent of the ``peacock patterns'' observed in \cite{Gu:2021ize}, where the possibility of a relation to BPS/DT invariants of local CY threefolds is also contemplated. (Also see Figures \ref{Fig:ray-diagrams-4d},  \ref{fig:F0-raydiagram}).

For local $dP_3$ we choose generators  $\gamma_i$ of the charge lattice corresponding to nodes of the BPS quivers shown in  Figure~\ref{Fig:dP3-intro}.
In a chamber where
\be
\begin{split}	
	& Z_{\gamma_1} = Z_{\gamma_4} \,,
	\quad
	Z_{\gamma_2} = Z_{\gamma_5} \,,
	\quad
	Z_{\gamma_3} = Z_{\gamma_6} \,,
	\\
	& \arg Z_{\gamma_1} > \arg Z_{\gamma_2}> \arg Z_{\gamma_3}\,,
	\quad
	Z_{\gamma_1}+Z_{\gamma_3} = Z_{\gamma_2 }\in \IR^+ \,.
\end{split}
\ee
we find the following BPS spectrum 
\begin{equation}
\begin{array}{|c|c|}
	\hline
	\gamma & \Omega(\gamma;y) \\
	\hline\hline
	\gamma_r + k (\gamma_1+\gamma_2+\gamma_3) & 1\\
	-\gamma_r + (k+1) (\gamma_1+\gamma_2+\gamma_3) & 1\\
	\gamma_s + k (\gamma_4+\gamma_5+\gamma_6) & 1\\
	-\gamma_s + (k+1) (\gamma_4+\gamma_5+\gamma_6) & 1\\
	\hline
	\gamma_t +k\gamma_{D0} & 1 \\
	-\gamma_t +(k+1)\gamma_{D0} & 1 \\
	\gamma_a+\gamma_b +k\gamma_{D0} & 1 \\
	-\gamma_a-\gamma_b +(k+1)\gamma_{D0} & 1 \\
	\gamma_1+\gamma_2 + \gamma_3+k\gamma_{D0} & y+y^{-1} \\
	-\gamma_1-\gamma_2-\gamma_3+(k+1)\gamma_{D0} & y+y^{-1} \\
	(k+1)\gamma_{D0} & y^3 + 4y+y^{-1}\\
	\hline
\end{array}
\end{equation}
with $k\geq  0$ and
\be
	r\in\{1,3\}
	\qquad 
	s\in\{4,6\}
	\qquad 
	t\in\{2,5\}
	\qquad
	(a,b) \in  \{(1,3), (4,6)\}\,.
\ee
where $\gamma_{D0} = \sum_{i=1}^6 \gamma_i$,  and again the spectrum  also  includes the respective antiparticles obtained by sending $\gamma\to-\gamma$. This spectrum consists of two copies of the weakly coupled spectrum of $SU(2)$ SQCD with one fundamental hypermultiplet in 4d \cite{Alim:2011kw}, with the extra KK towers in lower part of the table. This is quite striking, since the low-energy gauge theory phase of this SCFT is five-dimensional $N_f=2$ $SU(2)$ gauge theory.
As for local $\IF_0$, we observe that the spectrum  is organized into towers, over $\gamma_r,\gamma_s$ and $\gamma_t, \gamma_a+\gamma_b, \gamma_1+\gamma_2+\gamma_3, \gamma_{D0}$ with steps $\gamma_1+\gamma_2+\gamma_3, \gamma_4+\gamma_5+\gamma_6$ and $\gamma_{D0}$. This structure is reminiscent of the ``peacock patterns'' observed in \cite{Gu:2021ize}.

For local $dP_5$ we choose generators  $\gamma_i$ of the charge lattice corresponding to nodes of the BPS quivers shown in  Figure~\ref{Fig:dP5-intro}.
In a chamber where
\be
\begin{split}
	&Z_{\gamma_2}=Z_{\gamma_6}=c\cdot Z_{\gamma_1}=c\cdot Z_{\gamma_5}\,,
	\quad
	Z_{\gamma_4}=Z_{\gamma_8}=c\cdot Z_{\gamma_3}=c\cdot Z_{\gamma_7}\,,
	\\
	&
	Z_{\gamma_1}=\overline{Z}_{\gamma_3}\,,
	\quad
	\arg Z_{\gamma_1} > \arg Z_{\gamma_3}\,,
\end{split}
\ee
with an arbitrary real constant $c>1$, needed to ensure that $Z_{\alpha_i}$ belong to the positive half-plane,
we find the following BPS spectrum 
\begin{equation}
\begin{array}{|c|c|}
	\hline
	\gamma & \Omega(\gamma;y) \\
	\hline\hline
	\gamma_r + k (\gamma_1+\gamma_2+\gamma_3+\gamma_4) & 1\\
	-\gamma_r + (k+1) (\gamma_1+\gamma_2+\gamma_3+\gamma_4) & 1\\
	\gamma_s + k (\gamma_5+\gamma_6+\gamma_7+\gamma_8) & 1\\
	-\gamma_s + (k+1) (\gamma_5+\gamma_6+\gamma_7+\gamma_8) & 1\\
	\hline
	\gamma_a+\gamma_b + k \gamma_{D0} & 1\\
	-\gamma_a-\gamma_b + (k+1) \gamma_{D0} & 1\\
	\gamma_1+\gamma_2+\gamma_3+\gamma_4 + k \gamma_{D0} & y+y^{-1}\\
	-\gamma_1-\gamma_2-\gamma_3-\gamma_4 + (k+1) \gamma_{D0} & y+y^{-1}\\
	(k+1)\gamma_{D0} & y^3 + 6y+y^{-1}\\
	\hline
\end{array}
\end{equation}
with $k\geq  0$ and 
\be
\begin{split}
	& r\in \{1,2,3,4\},\quad s\in \{5,6,7,8\}  \\
	& (a,b)\in \{(1,3), (1,4), (2,3), (2,4), (5,7), (5,8), (6,7), (6,8)\} 
	\\
\end{split}
\ee
where $\gamma_{D0} = \sum_{i=1}^8 \gamma_i$,  and again the spectrum  also  includes the respective antiparticles obtained by sending $\gamma\to-\gamma$. This spectrum consists of two copies of the weakly coupled spectrum of $SU(2)$ SQCD with two fundamental hypermultiplets in 4d \cite{Alim:2011kw}, with the extra KK towers in lower part of the table. Again, this differs from what could be expected from the low-energy gauge theory phase of this SCFT, which is five-dimensional $N_f=4$ $SU(2)$ gauge theory.
As for local $\IF_0$ and $dP_3$, we observe once again that the spectrum  is organized into towers, over $\gamma_r,\gamma_s$ and $\gamma_t, \gamma_a+\gamma_b, \gamma_1+\gamma_2+\gamma_3+\gamma_4, \gamma_{D0}$ with steps $\gamma_1+\gamma_2+\gamma_3+\gamma_4, \gamma_5+\gamma_6+\gamma_7+\gamma_8$ and $\gamma_{D0}$, in a way that is reminiscent of ``peacock patterns''  \cite{Gu:2021ize}.

A byproduct of these results are new and highly nontrivial wall-crossing identities.
For example, for local $\mathbb{F}_0$ we find
\be
\begin{split}
	& \prod_{k\geq 0}^{\nearrow}\Phi( X_{\gamma_1 + k (\gamma_1+\gamma_2)}) \Phi( X_{\gamma_3+ k (\gamma_3+\gamma_4)}) 
	\\
	 &\qquad  \times
	\prod_{s=\pm1}\prod_{k\geq 0}\Phi((-y)^{s} X_{\gamma_1+\gamma_2+ k \gamma_{D0}})^{-1}
	\cdot \Phi((-y)^{s} X_{\gamma_3+\gamma_4+ k \gamma_{D0}})^{-1}
	\\
	 & \qquad  \qquad \times \prod_{k\geq 0}^{\searrow}\Phi( X_{\gamma_2 + k (\gamma_1+\gamma_2)}) \Phi( X_{\gamma_4+ k (\gamma_3+\gamma_4)}) 
	\\
	= &  \prod_{k\geq 0}^{\nearrow}\Phi( X_{\gamma_2 + k (\gamma_2+\gamma_3)}) \Phi( X_{\gamma_4+ k (\gamma_4+\gamma_1)}) 
	\\
	&\qquad \times 
	\prod_{s=\pm1}\prod_{k\geq 0}\Phi((-y)^{s} X_{\gamma_4+\gamma_1+ k \gamma_{D0}})^{-1}
	\cdot \Phi((-y)^{s} X_{\gamma_2+\gamma_3+ k \gamma_{D0}})^{-1}
	\\
	&\qquad \qquad\times \prod_{k\geq 0}^{\searrow}\Phi( X_{\gamma_3 + k (\gamma_2+\gamma_3)}) \Phi( X_{\gamma_1+ k (\gamma_4+\gamma_1)}).
	\\
\end{split}\label{eq:WCIF0}
\ee
Here 
\begin{align}
\Phi(x) = (-yx;y^2)_\infty^{-1}, && (x;y)_{\infty}\equiv\prod_{n=1}^{\infty}\left(1-y^nx \right),
\end{align}
and $\gamma_1\dots,\gamma_4$ are generators of the charge  lattice $\Gamma\simeq \IZ^4$, $X_\gamma$ are quantum-torus algebra variables with  product twisted  by the  DSZ pairing $X_\gamma X_{\gamma'} = y^{\langle\gamma,\gamma'\rangle}  X_{\gamma+\gamma'}$, and $\nearrow$ ($\searrow$) denotes increasing (decreasing) values of $k$ from left to right. The analogues of this identity for the case of $dP_3$, $dP_5$ are contained in equations \eqref{eq:WCIDP3} and \eqref{eq:WCIDP5} respectively.

\subsection*{Contents of the paper}

This paper is structured as follows. We introduce the necessary background in Section~\ref{sec:symmetry-chambers}; after recalling the notion of BPS quiver and mutation, we define what is meant by a collimation chamber, giving the example of pure $SU(2)$ super Yang-Mills and $\mathcal{N}=2^*$ theory in 4d. After this, we introduce the KS invariant $\IU$, and discuss how it can be constrained by using permutation symmetries of the quiver. In section \ref{sec:RootLats} we review how the lattice of BPS charges can be identified with the even-dimensional cohomology of the local Calabi-Yau, and for rank-1 SCFTs the sublattice of flavor charges is organized as an affine root lattice of type $E_n^{(1)}$. We then discuss how to extend to the whole BPS charge lattice of the local Calabi-Yau the action of the Cremona isometries of the del Pezzo surface, which are famously related to an action of the affine Weyl group on the $E_n^{(1)}$ root lattice. We identify this isometry group with the group of self-dualities of the five-dimensional theory, and formulate in section \ref{Sec:StabRoots} a set of criteria to associate a collimation chamber to an affine Weyl translation. In section \ref{sec:examples} we test our ideas in three specific examples, that of local $\mathbb{F}_0$, local $dP_3$ and local $dP_5$, for all of which we find appropriate collimation chambers and conjectural expressions for the wall-crossing invariants, leading to novel wall-crossing identities.

\section*{Acknowledgements}
We thank Sibasish Banerjee, Giulio Bonelli, Michele Cirafici, Cyril Closset, Michele Del Zotto, Alba Grassi, Boris Pioline, Mauricio Romo and Alessandro Tanzini for discussions and correspondence.
The work of FDM is supported by the NSERC through a CRM-ISM postdoctoral fellowship.
The work of PL is supported by NCCR SwissMAP, funded by the Swiss National Science Foundation.

\section{Symmetry constraints on wall-crossing invariants}\label{sec:symmetry-chambers}

\subsection{Quivers, stability conditions and BPS states}\label{sec:Quiver}

The notion of moduli space of stability conditions is a fundamental piece of data in the description of the spectrum of BPS states, in  the context of SUSY gauge theories, string theory, and mathematics.
In this paper we focus mainly on BPS states described by quivers, for which a notion of stability condition is provided by King's $\theta$-stability \cite{King1994MODULIOR}.\footnote{All quivers considered in this paper arise in the study of D-branes on local Calabi-Yau threefolds. See e.g. \cite{2002math.....12237B, 2016arXiv160300416B, Douglas:2002fj, 2000math......9209D, Douglas:2000gi} for further work on the relation to $\Pi$-stability and Bridgeland stability.}

Recall that a quiver $Q$ is an oriented graph, consisting of a set of $m$ vertices $Q_0$ and a set of oriented edges $Q_1$.
In what follows we will restrict to quivers without loops (arrows from a node to itself) or 2-cycles (i.e. where a pair of nodes is  connected by at least one arrow in each direction).
In this case the structure of $(Q_0,Q_1)$ may be encoded by an antisymmetric adjacency matrix $B_{ij}=-B_{ji}$ whose non-negative entries are the number of arrows from node $j$ to node $i$.\footnote{This convention follows \cite{Alim:2011ae} and other papers in the physics literature. It is opposite to  the convention often used in mathematics, where $B_{ij}>0$ would encode the number of arrows  $i\to j$.}
A quiver representation of dimension $\vec d=(d_1,\dots, d_{m})$ is a collection of vector spaces with dimension $\dim V_{i} = d_i$, and linear maps $\phi_a \in {\rm Hom}(V_i, V_j)$ for each arrow $a:i\to j$.
The stability data for the representation $R_{\vec d} = (\{V_{i}\},\{\phi_a\})$ is a collection of reals $\theta_i$ for each $i\in Q_0$. 
The role of $\theta_i$ in physics is that of Fayet-Iliopoulous couplings for an $\CN=4$ quantum mechanics associated to $R_{\vec d}$ in  the Higgs branch regime, see \cite{Douglas:1996sw, Denef:2002ru, Alim:2011ae} for details.
This quantum mechanics descibes the worldvolume dynamics of a collection of fractional branes with charges $\gamma_i$ for $i=1,\dots, m$. 
Let $\Gamma=\oplus_i\gamma_i\IZ$ be the charge lattice generated by nodes of the  quiver, and recall the  central charge is a linear function on $\Gamma$
\be\label{eq:Z-hom}
	Z\in {\rm Hom}(\Gamma,\IC)\,.
\ee
In a regime when central charges are nearly aligned, the FI couplings are related to the central charges as 
\be\label{eq:Z-to-theta}
	\theta_i = |Z_{\gamma_i}| \(\arg Z_{\gamma_i} - \arg Z_{\gamma}\)
\ee
where  $\gamma=\sum_i d_i\gamma_i$ is  the total charge of $R_{\vec d}$, and $Z_\gamma\in \IC$ is the central charge of a BPS state with charge $\gamma$.
The Dirac-Schwinger-Zwanziger pairing of basic  charges ie encoded by the quiver adjacency matrix
\be\label{eq:DSZ}
	\langle\gamma_i,\gamma_j\rangle = B_{ij} \,.
\ee
Thanks to linearity of central charge as a function of $\gamma$, the relation (\ref{eq:Z-to-theta}) allows to trade $\theta_i$ for  $Z_{\gamma_i}$ in discussing stability conditions. 

The values of FI couplings $\theta_i$, or equivalently central charges $Z_{\gamma_i}$, determine whether a representation $R_{\vec d}$ is (semi-)stable or not. 
In the case when (semi-)stable representations with given  dimension $\vec d$ exist, we denote by $\Omega(\gamma,y)$ the shifted Poincar\'e polynomial of the moduli space of such representations modulo gauge equivalence, where $\gamma=\sum_i d_i\gamma_i$.\footnote{The Poincar\'e polynomial is defined with respect to compactly supported de Rham cohomology of the moduli space $\CM_{\vec d}(\vec\theta)$ of (semi-)stable quiver representations. This is the appropriate definition for counting BPS states in string theory, but differs from the one that would be appropriate to discuss BPS states in  the  geometrically engineered gauge theory. For the latter, the appropriate notion is $L^2$ cohomology, which  has the property of respecting Poincar\'e symmetry. See \cite{Yi:1997eg, Duan:2020qjy, Mozgovoy:2020has} for details.}
In physics $\Omega(\gamma,y)$ coincides with (an uplift of) the Protected Spin Character introduced in \cite{Gaiotto:2010be}, which counts BPS states with different spin.
A convenient pictorial description of the BPS spectrum for a given choice of stability data is provided by \emph{ray diagrams}, where each BPS state is represented by a vector in the $Z$-plane, see Figure \ref{Fig:ray-diagrams-4d}. 

\paragraph{Mutations.}
Sometimes it is useful to consider how  the  quiver description of BPS states changes with different choices of positive half-plane (corresponding to ${\rm Re} Z>0$ in Figure~\ref{Fig:ray-diagrams-4d}).
If we keep the stability condition fixed, the BPS rays do not move. However the ray diagram changes whenever the choice of half-plane is tilted so that one of the rays exits from one side or the other (the  CPT-conjugate ray with $Z_\gamma\to-Z_\gamma$ enters on the other side). 
The quiver description of the BPS spectrum also changes, by a left or right mutation of $Q$ \cite{Berenstein:2002fi, Alim:2011ae, Alim:2011kw}.

Note that the BPS rays which are closest to either half-boundary of the positive $Z$ half-plane  always correspond to central charges of some of the quiver nodes.
When the half-plane is tilted clockwise, so that  a single ray exists, the quiver changes by a mutation on the node corresponding to the exiting ray.
A mutation $\mu_k$, acting on the node $k$ of $Q$, produces a new BPS quiver $Q'$ with adjacency matrix $B'_{ij}$ defined as follows
\begin{equation}\label{eq:BMutation}
B'_{ij}= \begin{cases}
-B_{ij}, & i=k\text{ or }j=k, \\
B_{ij}+\frac{B_{ik}|B_{kj}|+B_{kj}|B_{ik}|}{2},
\end{cases}
\end{equation}
At the same time, the charge vectors $\{\gamma_i\}$ labeling nodes of $Q$ also change to new labels $\{\gamma'_i\}$ for the nodes of $Q'$ as follows
\begin{equation}\label{eq:LeftMutations}
\mu_k(\gamma_j)= \begin{cases}
-\gamma_j, & j=k, \\
\gamma_j+[B_{jk}]_+\, \gamma_k, & \text{otherwise}.
\end{cases}
\end{equation}
If the half-plane is instead rotated counterclockwise, the mutation takes the form
\begin{equation}\label{eq:RightMutations}
\widetilde{\mu_k}(\gamma_j)= \begin{cases}
-\gamma_j, & j=k, \\
\gamma_j+[-B_{jk}]_+ \,\gamma_k, & \text{otherwise}.
\end{cases}
\end{equation}
The simultaneous changes in $B_{ij}$  and labels of nodes are compatible, they  preserve the property (\ref{eq:DSZ}).

\subsection{Collimation chambers}\label{sec:Coll}

In this work we focus on a specific class of stability conditions, which we name \emph{collimation chambers}.
The essential property of collimation chambers is that central charges of BPS states only accumulate along $\IR^{\pm}e^{i\vartheta_0}$, if at all.
For quivers of toric CY threefolds the real ray $\vartheta_0=0$ is canonically an accumulation ray, since the BPS spectrum always features towers of D0 branes with central charge $Z_{nD0} =  n\, 2\pi/R$ for $n\in \IZ\setminus\{0\}$, where $R$ is the radius of the M-theory circle.
An equivalent definition of collimation chambers, at least in all examples we considered, is that these are stability conditions for which the only BPS states with $Z_\gamma\notin \IR$ are hypermultiplets, i.e. their BPS index is $\Omega(\gamma)=1$.

\begin{figure}
\begin{center}
\includegraphics[width=.25\textwidth]{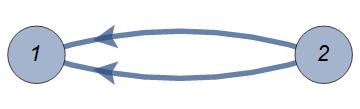}
\caption{$K_2$: BPS quiver for 4d pure $SU(2)$ super Yang-Mills. 
}
\label{fig:K2-quiver}
\end{center}
\end{figure}

The most basic example of a collimation chamber with at least one accumulation ray appears for the Kronecker quiver $K_2$ with $\langle\gamma_1,\gamma_2\rangle=2$ shown in Figure \ref{fig:K2-quiver}.
For stability conditions within the  chamber
\be
	\CC_1[K_2] : \quad \arg Z_{\gamma_2}>\arg Z_{\gamma_1}
\ee 
the ray diagram is the one of Figure \ref{Fig:raydiag-SW}.  This is the celebrated weak-coupling spectrum of 4d $\CN=2$ super Yang-Mills theory obtained by Seiberg and Witten \cite{Seiberg:1994rs}.

\begin{figure}
\begin{center}
\includegraphics[width=.25\textwidth]{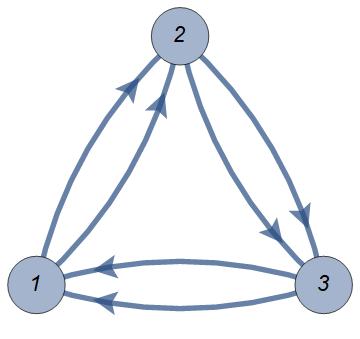}
\caption{$M_2$: BPS quiver for the $SU(2)$ $\mathcal{N}=2^*$ theory  
}
\label{fig:Markov-quiver}
\end{center}
\end{figure}

\begin{figure}[h]
\begin{center}
\begin{subfigure}{.5\textwidth}
\centering
\includegraphics[width=\textwidth]{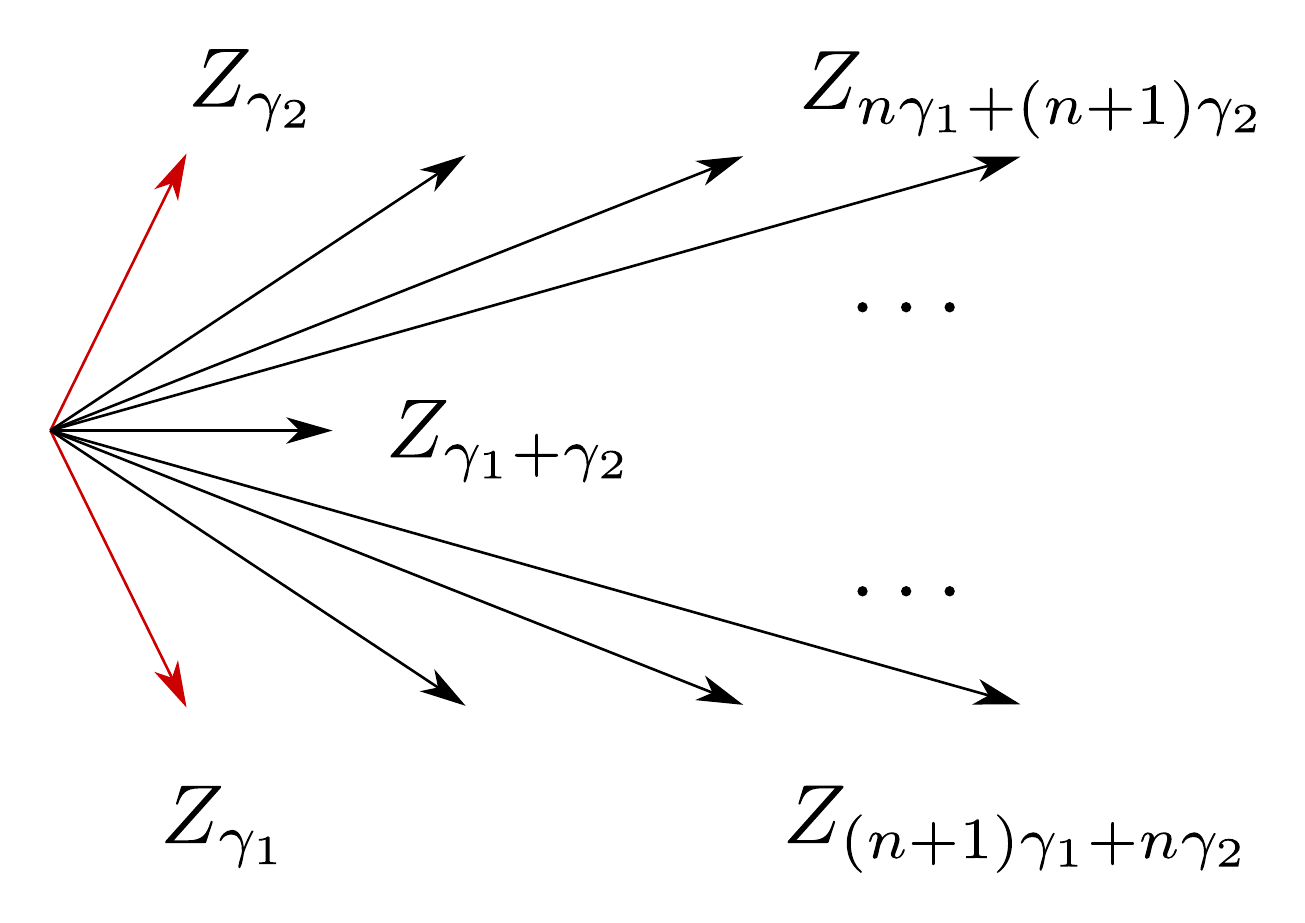}
\caption{Collimation chamber $\CC_1[K_2]$.}
\label{Fig:raydiag-SW}
\end{subfigure}\hfill
\begin{subfigure}{.5\textwidth}
\centering
\includegraphics[width=\textwidth]{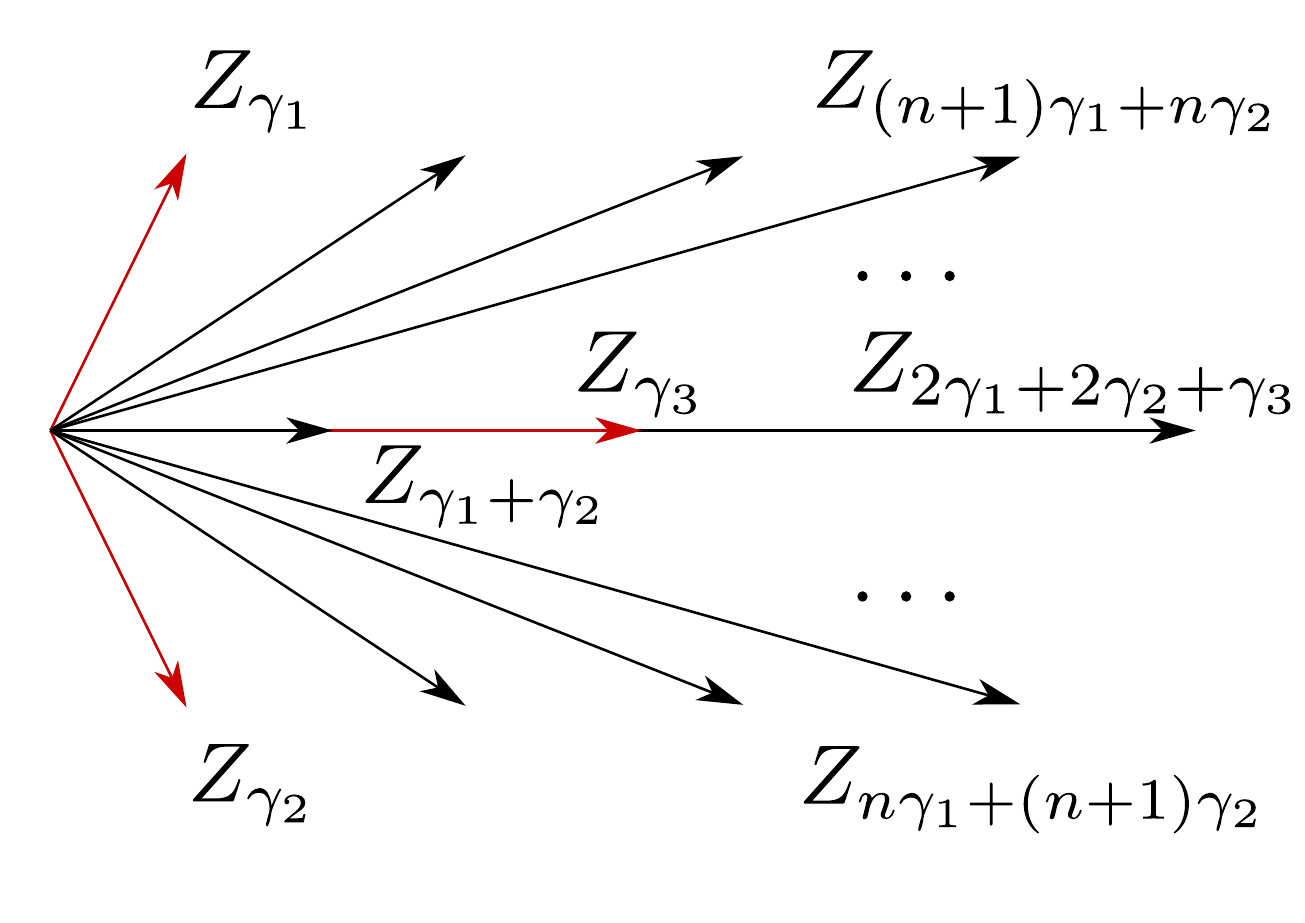}
\caption{Collimation chamber $\CC_1[M_2]$.}
\label{Fig:raydiag-N2star}
\end{subfigure}
\caption{Ray diagrams of collimation chambers.}
\label{Fig:ray-diagrams-4d}
\end{center}
\end{figure}

Another example of collimation chamber arises for the  Markov quiver $M_2$ with  $\langle\gamma_i,\gamma_{i-1}\rangle=2$ shown in Figure \ref{fig:Markov-quiver}.
The ray diagram for this chamber is shown in Figure \ref{Fig:raydiag-N2star}, and corresponds to the chamber
\be\label{eq:Markov-collimation-chamber}
	\CC_{1}[M_2]: \quad \arg Z_{\gamma_1} > \arg Z_{\gamma_3}=\arg (Z_{\gamma_1+\gamma_2}) > \arg Z_{\gamma_2} \,.
\ee
This chamber was studied in unpublished joint work by Greg Moore and the second-named author, and was used to obtain a closed-form description of the BPS spectrum of $\CN=2^*$ super Yang-Mills theory  \cite{Longhi:2015ivt,Longhi:2016wtv}.
The $M_2$ quiver has  two more collimation chambers  $\CC_{2}[M_2], \CC_{3}[M_2]$ related by  cyclic permutations of the central charges in (\ref{eq:Markov-collimation-chamber}).

The Markov example is especially interesting, because it turns out that the spectrum  of $\CN=2^*$ theory is generically very intricate, featuring infinitely many accumulation rays. 
The only known exception is precisely the collimation chamber of Figure \ref{Fig:raydiag-N2star} (or its $\IZ_3$ images $\CC_2,\CC_3$), where the BPS spectrum admits a simple exact description.
In fact this correlation between collimation chambers and nice BPS spectra turns out to generalize to (at least certain) quivers of toric Calabi-Yau threefolds, suggesting that closed-form exact descriptions of the spectrum can be obtained by studying such chambers. 
In the rest of this paper we explore and realize this idea for a few selected examples.

\subsection{The wall-crossing invariant $\IU$}\label{sec:wall-crossing-invariant}

The moduli space of stability conditions  has a wall-and-chamber structure, with different BPS spectra in different chambers.
Jumps of the spectrum are described by the wall-crossing formulae \cite{Kontsevich:2008fj, Joyce:2008pc}. We will  adopt here the  framework of Kontsevich and Soibelman, whose formula can be stated as a conservation law for a certain invariant $\IU$ built from the BPS spectrum.

Assume ${\rm Re} Z_{\gamma_i}>0$, $ \forall i\in  Q_0$, we shall refer to BPS particles as those BPS states whose central charge lies in the right-half of the $Z$-plane, and anti-particles as those in the other half\footnote{We make this choice of half-plane instead of the more traditional upper half-plane so as to have the D0-brane central charge, which is  real and positive, at  the center of the domain.}.
Let $\Gamma = \oplus_{i}  \gamma_i \, \IZ$ denote the lattice of charges generated by quiver nodes, and consider  the quantum torus algebra of formal variables $X_\gamma$ obeying
\be
	X_{\gamma}X_{\gamma'} = y^{\langle\gamma,\gamma'\rangle} X_{\gamma+\gamma'} \,.
\ee
The Kontsevich-Soilbeman invariant of wall-crossing is a product of quantum dilogarithms
\be\label{eq:factorization}
	\IU = \prod_{\gamma\in\Gamma_+}^{\curvearrowleft}\prod_{m\in \IZ} \Phi((-y)^m X_{\gamma})^{a_m(\gamma)}
\ee
where
\begin{itemize}
\item $\Gamma_+ = \{\gamma\in \Gamma, {{\rm Re} Z_\gamma}>0\}$ 
\item the ordering $\curvearrowleft$ is by increasing $\arg(Z_\gamma) $ to the left
\item $\Omega(\gamma,y) = \sum_m (-y)^m a_{m}(\gamma)$ is the Protected Spin Character \cite{Gaiotto:2010be}, which recovers the BPS index for $y\to -1$. 
\item  $\Phi(x) = (-yx;y^2)_\infty^{-1} = \prod_{s\geq 0}(1+y^{2s+1} x)^{-1}$
\item $m\in \IZ$ is the eigenvalue of $2 J_3$, a Cartan generator of the rotation group in three dimensions\footnote{A standard physical realization of  BPS quivers we study below in Section (\ref{sec:examples}) is M-theory on a CY threefold $X$ times $S^1\times \IR^4$. In this setup the quiver describes the spectrum of M2 and M5 branes wrapping along $\IR\subset \IR^4$, breaking $\mathfrak{so}(4)$ to $\mathfrak{so}(3)$.} 
\end{itemize}

When the stability condition belongs to a collimation chamber, the wall-crossing invariant takes the form
\be\label{eq:collimation-factorization}
	\IU = \IU(\measuredangle^+)\cdot\IU(\IR^+)\cdot\IU(\measuredangle^-)
\ee
where $\measuredangle^+$ and $\measuredangle^-$ correspond respectively to angular sectors $0<\arg Z<\pi/2$ and $-\pi/2\leq \arg Z<0$. The terms $\IU(\measuredangle^\pm)$ contain only (possibly countably infinitely many) hypermultiplet factors $\Phi(X_\gamma)$. \footnote{This is based on  the alternative characterization of collimation chambers mentioned in Section \ref{sec:Coll}.} 
A consequence of this, is that in collimations chambers it is straightforward to compute $\IU(\measuredangle^\pm)$ by standard techniques based on quiver mutations \cite{Alim:2011ae, Alim:2011kw, Closset:2019juk}. 
The only nontrivial task is the computation of $\IU(\IR^+)$.

\begin{remark}

For the BPS spectrum to be well-defined in a  collimation chamber, one must ensure that the marginal stability  condition is not verified. 
This implies that for any pair of $\gamma,\gamma'\in\Gamma$ such that $Z_\gamma//Z_{\gamma'}$ (either on the real ray, or away from it) their pairing vanishes $\langle\gamma,\gamma'\rangle=0$. 
However, this requirement could also be slightly relaxed. For example, if one is willing to forgo a factorization of $\IU(\IR^+)$ into quantum dilogarithms, then one may still make sense of this factor as a formal series in $X_\gamma$ even if the states with real central charge are at marginal stability with respect to each other.

\end{remark}

\subsection{Quiver permutation symmetry}\label{sec:Q-permutation-symmetry}
Before proceeding with  the task  of determining $\IU(\IR^+)$, we take a  brief technical detour to introduce the notion of permutation symmetries of a quiver. 

Let $\Pi_Q$ be the group of permutations acting on the labels of $Q_0$ which preserves the  structure of the quiver.
This is the subgroup of $S_{|Q_0|}$ of elements 
\be
	\pi : \gamma_i\mapsto\gamma_i' = \gamma_{\pi(i)}
\ee
such that the quiver $(Q,\{\gamma'_i\})$ is identical to the original one $(Q,\{\gamma_i\})$.
For example the $M_2$ quiver of Figure \ref{fig:Markov-quiver} has a permutation  symmetry $\Pi_{M_2}\simeq \IZ_3$, acting as $\gamma_i\mapsto \gamma_{i+1 \,{\rm mod}\, 3}$. An example of a quiver with  trivial symmetry group is the Kronecker quiver (\ref{fig:K2-quiver}).

Later we will encounter a more general  definition of quiver  symmetries, involving  mutations. The group of $\Pi_Q$ is  a subgroup of the full automorphism group ${\rm Aut}_Q$ of the quiver,  whose  definition will be introduced in section \ref{sec:MutFlows}). 
For later convenience, we highlight a few important features of mutations and permutations. Here we simply state these as (partly obvious) facts, their scope will become more clear with  examples below.
\begin{itemize}
\item A mutation cannot change the stability condition. A permutation can.
The action of $\pi\in \Pi_Q$ restricts to permutations of collimation chambers $\CC_\alpha\mapsto\CC_{\pi(\alpha)}$. 
This  action is generically not faithful. For an example we refer to Observation \ref{obs:F0-1} for local  $\IF_0$.
\item Stability conditions related by a permutation are always connected by a continuous path along which no BPS ray  exits the positive  half-plane of central  charges (in our case, this is the half-plane ${\rm Re} Z_\gamma\ge0$). 
For an example see Observation \ref{obs:F0-2} for local  $\IF_0$.
\item A mutation changes the choice of positive half-plane of central charges by tilting it (until the BPS ray of the mutant note exists the half-plane \cite{Alim:2011ae}). A permutation does not.
\item The group $\Pi_Q$ isn't preserved by mutations, because mutations generically change the  structure of a quiver, hence its automorphism group. We will work with  a fixed representative $Q$ in the mutation class $[Q]$, and consider its symmetry group  $\Pi_Q$. 
\end{itemize}

\subsection{Constraining $\IU$}\label{sec:U-constrained}

The outstanding task of computing $\IU(\IR^+)$ in a collimation chamber can be achieved by taking advantage of the permutation symmetry of a quiver $Q$, whenever $\Pi_Q$ is nontrivial.
A prototypical example is the $\IZ_4$ symmetry of the BPS quiver of local $\IF_0$ in Figure~\ref{Fig:QuiverF0}.  In this case, it was shown in \cite{Longhi:2021qvz} that constraints from this symmetry are strong enough to determine $\IU(\IR^+)$.\footnote{More precisely, up  to factors involving  pure-flavor  states, such as $D0$   branes in this case. The reason is that wall-crossing invariance doesn't constrain these spectrum of purer-flavor states.}
Here we review the basic idea, and explain how it its generalizes to other quivers with nontrivial symmetries.

When a labeled quiver $(Q,\{\gamma_i\})$ has a nontrivial permutation symmetry $\Pi_Q$, this imposes constraints on the wall-crossing invariant $\IU$.
To illustrate this, let us write the  latter as a formal series in $X_\gamma$
\be
	\IU = \sum_{\gamma\in\Gamma_+} u_\gamma X_\gamma
\ee
where coefficients $u_\gamma$ are Laurent polynomials in $y$ with integer coefficients.
Acting with $\pi$ induces a permutation of the $\gamma_i$, which induces a  relabeling of formal variables $X_\gamma$
\be\label{eq:symmetry-derivation}
	\IU 
	\quad\mathop{\longrightarrow}^{\pi}\quad
	\IU' = \sum_{\gamma} u_{\gamma} X_{\pi(\gamma)} = \sum_{\gamma} u_{\pi^{-1}(\gamma)} X_{\gamma}
	= \sum_{\gamma} u'_{\gamma} X_{\gamma}
\ee
where we used the fact that $\gamma\in  \Gamma_+ \leftrightarrow \pi(\gamma)\in \Gamma_+$.\footnote{Note that $\Gamma_+=\{\gamma\in \Gamma| \ \gamma = \sum_{i=1}^{Q_0}n_i \gamma_i ,\  n_i\geq 0\}$. By definition $\Gamma_+$ is the subset of charges with ${\rm Re} Z_\gamma>0$, which is  the quiver half-plane. It follows  that  any such $\gamma$ is a non-negative integer linear combination of the charges $\gamma_i$ associated to the vertices of $Q$.}

Next, recall  from Section \ref{sec:Q-permutation-symmetry} that the symmetry $\pi$ may change the stability condition by permuting the central charges associated to the nodes, but the new configuration of central charges is connected to the old one by a smooth motion in the moduli space of $\{Z_{\gamma_i}\}\ \simeq \IC^{Q_0}$ throughout which no rays exit or enter the half-plane ${\rm Re Z}>0$. This implies that $\IU'=\IU$ by  wall-crossing invariance, and leads to the non-trivial  constraint
\be\label{eq:symmetry-equation}
	u_{\pi(\gamma)}  = u_\gamma\,.
\ee

Taken alone, these constraints are generically too weak to fully determine $\IU$ itself. But  in  a collimation chamber we can  factorize $\IU$ as in (\ref{eq:collimation-factorization}) into $\IU(\measuredangle^\pm)$ and $\IU(\IR^+)$.
As mentioned previously, it is often easy to  compute $\IU(\measuredangle^\pm)$ by standard techniques such as the mutation algorithm. With this input,  (\ref{eq:symmetry-equation}) becomes powerful enough to determine $\IU(\IR^+)$ entirely, or almost entirely (details below). If we further require  $\IU(\IR^+)$  to factorize into quantum  dilogarithms, we typically  find that symmetries determine the exponents $\Omega(\gamma;y)$  uniquely.

Concretely, we shall often adopt a formal  series expression for $\IU(\IR^+)$
\be\label{U-real-expansion}
	\IU(\IR^+) = \sum_{\gamma\in \Gamma_{\IR^+}} c_\gamma \, X_\gamma
\ee
where $ \Gamma_{\IR^+} =  \{\gamma\in \Gamma | \ Z_\gamma\in \IR^+\}$ and $c_\gamma \in \IZ[y,y^{-1}]$.
We then compute the coefficients $u_\gamma(y)$ of $\IU$ as a function of $c_\gamma(y)$ through (\ref{eq:collimation-factorization}), and finally impose the  symmetry equations (\ref{eq:symmetry-equation}) on the $u_\gamma(y)$ to fix the  coefficients $c_\gamma(y)$ of $\IU(\IR^+)$.   
We will illustrate this in Section \ref{sec:examples} with several examples.

\section{Affine root lattices and stability conditions}\label{sec:RootLats}

In  the previous section we discussed collimation chambers $\CC_\alpha$,  and explained how they can be used in conjunction with quiver permutation symmetries $\Pi_Q$ to obtain  equations for $\IU$ which  determine  the  invariant. 
Here we discuss in more detail the larger automorphism group $\Aut_Q$ of a quiver $Q$, of which  permutations are generically a subgroup
\be
	\Pi_Q \subset \Aut_Q\,.
\ee  
In the cases of interests to this paper, $\Aut_Q$ contains a subgroup isomorphic to the extended affine Weyl group\footnote{This is defined as the semidirect product of the affine Weyl group of the Lie algebra and its outer automoprhisms.} for the affine Lie algebra associated to $Q$ \cite[Theorem 3.1]{Bershtein:2017swf} (also see \cite{mizuno2020q} for a comprehensive discussion). 
The affine root lattice can be naturally embedded within the charge lattice $\Gamma$ \cite{Bonelli:2020dcp}.
One of the main points of this section is to illustrate how 
the action of affine Weyl translations on $\Gamma$ can be used, at least in some cases, to define stability conditions corresponding to collimation chambers.

\subsection{Charge lattices}\label{sec:charge-geom}

Affine root lattices naturally describe the flavor symmetry group of five-dimensional SCFTs, from the geometry of the Calabi-Yau threefold from which the field theory is obtained in M-theory compactification \cite{Seiberg:1996bd,Intriligator:1997pq}. In the case of rank-1 theories, the relevant geometry consists of local del Pezzos 
$X = K_{S}$ with $S=dP_n$
$n=1,\dots,8$, 
together with local $S=\IF_0$\footnote{More precisely, the local del Pezzo corresponds to a gauge theory phase, while its blowdown to the UV superconformal fixed point.}. 
These are local Calabi-Yau threefolds defined as total space of the canonical line bundle of a suface $S$,
obtained by blowing up $\mathbb{P}^2$ at $n$ points or $\mathbb{P}^1\times\mathbb{P}^1$ at $n-1$ points, $n=1,\dots,8$.
The cases with $S=dP_{0,1,2,3}$ are toric, while the others are not (although they still have many properties of toric manifolds, in particular shrinking one-cycles).

We now recall a few well-known facts about BPS states of type IIA string theory on $X$, see e.g.  \cite{Beaujard:2020sgs,  Mozgovoy:2020has} for recent accounts with additional details.
The physical moduli space of the theory is called the ``extended Coulomb branch'' \cite{Closset:2018bjz}, including Coulomb moduli, mass parameters, and the D0 brane mass.
This moduli space is geometrically realized by the K\"ahler parameters of the Calabi-Yau, parametrizing the resolution of its singularities and the volume of the only exceptional four-cycle, and by the radius of the  M-theory circle.\footnote{We slightly abuse terminology in  calling this  the ``extended Coulomb branch'', since in \cite{Closset:2018bjz} the M-theory circle radius is not included as a parameter of the theory. 
Rather the redius is just seen  as setting a scale, which is then used to multiply masses to produce  dimensionless K\"ahler parameters. 
In connection to 4d limits, form the viewpoint of field theory, one can consider various scalings where the radius is taken to zero, but they are equivalently described as limits in the Kahler moduli space. 
In a 5d theory on $S^1\times \IR^4$ however, where we are interested in the moduli space parametrizing BPS central charges, it makes sense to include the M-theory radius.
Its role is to set the overall scale of central charges, such a rescaling isn’t necessarily included in the parametrization by Kahler moduli, at least a priori.
We thank Cyril Closset for helpful comments on this point.
} 
BPS charges are Chern characters of compactly supported coherent sheaves, with values in the charge lattice\footnote{Actual charges would be defined by K-theory classes. The distinction between the two involves issues related to integrality of the  charge  lattice. For the purpose of this work we can safely ignore these.}
\be\label{eq:physical-lattice}
\Gamma := H^{{\rm even}}_{{\rm cpt}} = H^0_{}(X,\mathbb{Z})\oplus H^2_{cpt}(X,\mathbb{Z})\oplus H^4_{cpt}(X,\mathbb{Z})\,.
\ee
The lattice $\Gamma$ is endowed with a skew-symmetric bilinear form $\langle\,\cdot\,,\,\cdot\,\rangle$,  arising as the antisymmetrized Euler form. Namely, for sheaves $E_\gamma$ with charge $\gamma$, the pairing is
\be\label{eq:DSZ-euler}
	\langle E_\gamma,  E_{\gamma'}\rangle := \chi(E_\gamma,E_{\gamma'}) - \chi(E_{\gamma'},E_\gamma)
\ee
where
\be\label{eq:Euler-form}
	\chi(E_\gamma,E_{\gamma'}) = \int_{S} \ch(E_\gamma^\vee)\,\ch(E_{\gamma'})\,\Td(S)
\ee
where $E^\vee$ is the dual sheaf to $E$, and $\Td$ denotes the Todd characteristic class.
In gauge theory descriptions the pairing $\langle\,\cdot\,,\,\cdot\,\rangle$ agrees with the Dirac-Schwinger-Zwanziger pairing~(\ref{eq:DSZ}), encoded in the quiver structure. Quiver nodes correspond to a choice of exceptional collection for $S$. 
For $X=K_S$ with $S$ a del Pezzo surface, the DSZ pairing matrix has rank two. This induces a (noncanonical) splitting of the charge lattice into flavor and gauge  charges
\be
\Gamma \simeq \Gamma_f\oplus\Gamma_g,
\ee
locally on moduli space. The flavor sublattice is characterized  by the vanishing of $\langle\,\cdot\,,\,\cdot\,\rangle$ when  restricted to $\Gamma_f\simeq \IZ^{f}$. 
We also note that in all examples of this  class $H^4_{{\rm cpt}}  \simeq \IZ$ is a sublattice of $\Gamma_g\simeq \IZ^2$. 
When $S=dP_n$, we have $\dim H^2(S,\IZ) = n+1$. Since $\dim H^0(X,\IZ)= \dim H^4_{{\rm cpt}}(X,\IZ)=1$, these add up to $\rk \Gamma = n+3$. Moreover since $\rk\Gamma_g=2$, it follows that $\rk \Gamma_f=n+1$. 
Clearly all the interesting information about $\Gamma$ is contained in $H^2_{{\rm cpt}}(S,\IZ)$. We will discuss its structure in greater detail next.

\subsection{Affine symmetry from geometry}\label{sec:affine-sym} 

While $X$ is noncompact, by restricting to compactly supported sheaves we have  $H^2_{cpt}(X,\mathbb{Z})\simeq H^2_{}(S,\mathbb{Z})$, and furthermore we can invoke Poincar\'e duality to identify 
\begin{equation}
	H_2^{}(S,\mathbb{Z})\simeq H^2_{}(S,\mathbb{Z})=\Pic(S)\,.
\end{equation}
Under this identification, generators of $H_2^{cpt}(S,\mathbb{Z})$ are mapped to divisors of $S$ (i.e. line bundles) in $\Pic(S)$. 
As observed in \cite{Iqbal:2001ye, Hanany:2001py}, a natural basis of $H_2^{cpt}(S,\mathbb{Z})$ is provided by
homology classes of the exceptional curves $\mathcal{L}_1,\dots,\mathcal{L}_n$ obtained by blowing up points in $\mathbb{P}^2$, as well as the pull-back of the hyperplane $\mathcal{L}_0\in \text{Pic}(\mathbb{P}^2,\mathbb{Z})$. If we view instead $dP_n$ as obtained by blowups on $\IF_0$, for $n\ge 1$ we can consider the basis of $\text{Pic}(dP_n)$ given by $\mathcal{H}_1,\mathcal{H}_2$ (which are total transforms of lines in $\IF_0$), and $\mathcal{E}_1,\dots,\mathcal{E}_{n-1}$, which are the exceptional classes of the blown-up points. Then
\begin{equation}
\text{Pic}(S)=\bigoplus_{i=0}^n\mathcal{L}_i=\(\mathcal{H}_1\oplus\mathcal{H}_2\)\oplus \bigoplus_{i=1}^{n-1}\mathcal{E}_i.
\end{equation}
The intersection form between the cycles is
\begin{align}\label{eq:L-pairing}
	\chi\left(\mathcal{L}_0 ,  \mathcal{L}_0 \right)=1, && \chi\left(\mathcal{L}_0, \mathcal{L}_i \right)=0, && \chi\left(\mathcal{L}_i, \mathcal{L}_j \right)=-\delta_{ij}, && i,j=1,\dots, n,
\end{align}
\begin{align}\label{eq:HE-pairing}
	\chi\left(\mathcal{H}_m, \mathcal{H}_k \right)=1-\delta_{km}, && \chi\left(\mathcal{H}_m,\mathcal{E}_k \right)=0, && \chi\left(\mathcal{E}_m, \mathcal{E}_k \right)=-\delta_{mk},
\end{align}
and the anti canonical class is 
\begin{equation}\label{eq:AnticanClass}
	\mathcal{K}_S =3\mathcal{L}_0-\mathcal{L}_1 \dots-\mathcal{L}_n=2\mathcal{H}_1+2\mathcal{H}_2-\mathcal{E}_1 \dots-\mathcal{E}_{n-1}.
\end{equation}
A remarkable fact about these geometries \cite{Iqbal:2001ye, Hanany:2001py} is that there exist a subset of $n$ generators $\{\alpha_i\}$ of $\Pic(dP_n)$
for which the intersection form is (minus) the Cartan matrix of the $E_n$ algebra:
\begin{equation}
\left(\alpha_i,\alpha_j \right)=-C_{ij}^{(E_n)},
\end{equation}
so that $\{\alpha_i\}$ can be regarded as the simple roots of $E_n$.

Together with the only generator $\delta$ of $H^0(X,\IZ)$, with which they all have vanishing pairing (\ref{eq:Euler-form}) suitably extended  to $X$, they generate the flavor lattice $\Gamma_f$. Physically $\delta$ is identified with the D0 brane charge. 
This is therefore isomorphic to the root lattice $\CQ(\widehat{\mathfrak{g}})$ of the  affine Lie algebra $E_n^{(1)}$, where $\delta$ is the null root
\be\label{eq:flavor-roots}
	\Gamma_f \, \simeq\,  \CQ(\widehat{\mathfrak{g}})\,.
\ee
The gauge charge lattice $\Gamma_g\simeq\IZ^2$ is instead generated by the remaining cycle in $\Pic(S)$ and by the unique generator of $H^4_{{\rm cpt}}(X,\IZ)$.

\begin{remark} 
The DSZ pairing descends from the Euler form (\ref{eq:DSZ-euler}) by antisymmetrization, and it is trivial on $\Gamma_f$. 
The DSZ pairing plays a prominent role in the dynamics of BPS states, and their wall-crossing phenomena,  and encodes the structure of the BPS quiver $Q$. On the other hand, the information encoded by the Euler form, which is  hidden by the antisymmetrization, contains key information about the symmetries of the underlying theory  embedded in the structure of $\Gamma_f$  \cite{Iqbal:2001ye, Hanany:2001py}.
Remarkably, information about these symmetries is not quite lost. As we will review below, they are encoded in $\Aut_Q$.
\end{remark}

\begin{table}
\centering
\begin{tabular}{ c c c c c c c c}

$E_8^{(1)}$ & $E_7^{(1)}$ & $E_6^{(1)}$ & $E_5^{(1)}$ & $E_4^{(1)}$ & $E_3^{(1)}$ & $E_2^{(1)}$ & $E_1^{(1)}$ \\
\\
$E_8^{(1)}$ & $E_7^{(1)}$ & $E_6^{(1)}$ & $D_5^{(1)}$ & $A_4^{(1)}$ & $(A_2+A_1)^{(1)}$ & $(A_1+A_1)^{(1)}$ & $A_1^{(1)}$
\end{tabular}
\caption{Notation for affine exceptional algebras}
\end{table}

\subsection{Cremona  isometries, affine translations, and $\Aut_Q$}\label{sec:MutFlows}

The Cremona isometries $\Cr(S)$  automorphisms of $\Pic(S)$ that preserve the intersection form, the canonical class and the semi-group of effective divisors of $\Pic(S)$ \cite{2001CMaPh.220..165S, joshi2019discrete}.
Since they preserve the canonical class, they are also the automorphisms of the local del Pezzo CY manifolds (that are the total space of the canonical bundle), preserving the intersection form and the semi-group of effective divisors of the base.%
\footnote{%
As we have reviewed above, from the point of view of the local Calabi-Yau, in order to get the affine root lattice it is necessary to consider not just $Pic(S)$, but also $H^0(X)$, which contains the null root. However, the del Pezzo contains all the information about the local Calabi-Yau through its canonical divisor. Consequently, in the literature on q-Painlev\'e equations, where affine root lattices are traditionally seen as arising from del Pezzo surfaces, the null root is identified with the canonical divisor of $dP_9$ (note that equation \eqref{eq:AnticanClass} implies that $\chi(\mathcal{K}_{dP_9},\mathcal{K}_{dP_9})=0$), which is then mapped to lower del Pezzos through a blowdown procedure \cite{2001CMaPh.220..165S}.
}
It is well-known that such transformation are realized as the extended affine Weyl group $\widetilde{W}(\widehat{g})$ acting not only on  the root lattice $\Gamma_f$, but on the whole lattice $\Gamma$ \cite{GHK2015,mizuno2020q}.

It was shown in \cite{Bershtein:2017swf, mizuno2020q} that $\Aut_Q$ contains a subgroup isomorphic to the extended affine Weyl group $\widetilde{W}(\widehat{\mathfrak{g}})$ for all quivers of del Pezzo surfaces up to $dP_8$ and of $\IF_0$
\be\label{eq:Aut-W}
	\Aut_Q \supseteq \widetilde{W}(\widehat{\mathfrak{g}})\,.
\ee
Thus elements  $\Aut_Q$, which are realized by specific sequences of mutations, permutations and inversion, provide a realization of the Cremona group of $dP_n$ on the associated BPS charge lattice $\Gamma$.
Since $\Aut_Q$ preserves the form of the quiver, it may be regarded as the \emph{self-duality} group of the  quiver  quantum mechanics.

\begin{figure}[t]
\begin{center}
\begin{tikzpicture}[scale=.9]
\node at ($(0,0)$) {$E_8^{(1)}$};
\node at ($(2,0)$) {$E_7^{(1)}$};
\node at ($(4,0)$) {$E_6^{(1)}$};
\node at ($(6,0)$) {$E_5^{(1)}$}; 		
\node at ($(8,0)$) {$E_4^{(1)}$};
\node at ($(10,0)$) {$E_3^{(1)}$};
\node at ($(12,0)$) {$E_2^{(1)}$};
\node at ($(14,0)$) {$A_1^{(1)}$};
\node at ($(16,0)$) {$A_0^{(1)}$};
\node at ($(14,2)$) {$\begin{matrix}A_1^{(1)} \\|\alpha|^2=8\end{matrix}$};

\draw[thick,->](0.5,0) to (1.5,0);
\draw[thick,->](2.5,0) to (3.5,0);
\draw[thick,->](4.5,0) to (5.5,0);
\draw[thick,->](6.5,0) to (7.5,0); 		
\draw[thick,->](8.5,0) to (9.5,0);
\draw[thick,->](10.5,0) to (11.5,0);
\draw[thick,->](12.5,0) to (13.5,0);

\draw[thick,->](12.4,0.5) to (13.2,1.3);
\draw[thick,->](14.7,1.3) to (15.5,0.5);

\node (F0) at (14,-2) {\includegraphics[width=80pt]{figures/QuiverF0}};
\node (dP3) at (10,-2) {\includegraphics[width=80pt]{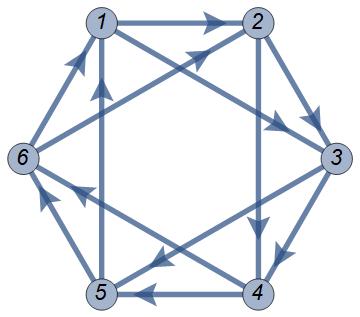}};
\node (dP5) at (6,-2) {\includegraphics[width=80pt]{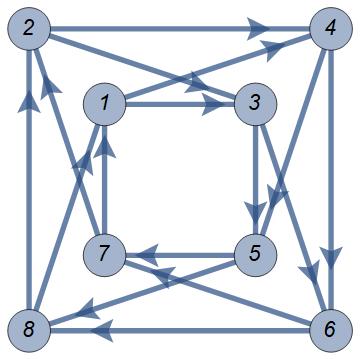}};

\end{tikzpicture}
\end{center}
\caption{Symmetry algebra of 5d SCFTs (or symmetry type of q-Painlev\'e equations) with BPS quivers for local $dP_5,dP_3,\mathbb{F}_0$.
}
\label{Fig:Sakai}
\end{figure}

When extending the representation of $\widetilde{W}(\widehat{g})$ from $\Gamma_f$, where it is naturally defined, to $\Gamma$, there are subtleties to deal with: in order to  have a proper representation of $\widetilde{W}(\widehat{g})$ on $\Gamma$, it seems necessary to use transformations that include both clockwise and anticklockwise mutations, as was done in \cite{mizuno2020q}.
However this is unnatural from a physical point of view: we require that the action must include only left (or only right) mutations, in keeping with our goal of connecting to tilting induced by  stability conditions as in  the mutation method of \cite{Berenstein:2002fi, Alim:2011ae, Alim:2011kw, Cecotti:2011rv}.
Instead, we will adopt a realization of $\widetilde{W}(\widehat{g})$ involving only left mutations as in \cite{Bershtein:2017swf}.\footnote{Clearly, it would be important to  determine to what extent the condition of using only left (or right) mutations determines \emph{uniquely} the realization of $\widetilde{W}(\widehat{g})\subset \Aut_Q$.}
This will result in an action on  $\Gamma$ which  is generally \emph{not} a
 proper representation of $\widetilde{W}(\widehat{g})$, but restricts to a proper representation on the root lattice $\Gamma_f$ (on $\Gamma$ we have a representation ``modulo seed isomorphisms'' \cite{GHK2015,mizuno2020q}).
Recall that affine Weyl groups have the following  decomposition into reflections and  translations
\be\label{eq:affine-translations}
	W(\widehat{\mathfrak{g}})  = W(\mathfrak{g})\ltimes \CT(\widehat{\mathfrak{g}}) \,.
\ee
In the context of discrete Painlev\'e equations the group $\CT(\widehat{g})$  corresponds to  the flow described by the equations themselves, while other elements of $\widetilde{W}(\widehat{g})$ act as symmetries of the equation. 
Following the general philosophy of \cite{Bonelli:2020dcp}, we want to study the action of affine translations  on the lattice $\Gamma$ of BPS charges, to study the BPS spectral problem. 
To this end, we recall the quiver description of BPS states reviewed in  Section \ref{sec:Quiver}.
For general toric geometries, BPS quivers can be determined by resorting to brane tiling techniques (see \cite{Franco:2017jeo} for a recent review). 
Other options include: local mirror symmetry \cite{Hanany:2001py}, 
studying calibrated one-cycles on the mirror curve 
\cite{Klemm:1996bj, Eager:2016yxd, Banerjee:2019apt, Banerjee:2020moh},
directly deriving the adjacency matrix directly from $\Pic(S)$ as in \cite{mizuno2020q}, and the approach based on the geometry of loop groups of \cite{Fock:2014ifa}.
We collect the relevant quivers in Figure  \ref{Fig:Sakai}.

Having identified the relevant BPS quivers (labeled by $\gamma_i$), the next step is to explain how these are related the affine Weyl group action on $\Gamma$.
Recall the definition of mutations from section \ref{sec:Quiver}. In addition to this we'll need another important transformation, called inversion and denoted by $\iota$. 
This operation reverses the orientation of all arrows in $Q$, and flips the signs of all label charges
\begin{align}
\iota(B)=-B, && \iota(\gamma_i)=-\gamma_i.
\end{align}
The automorphism group $\Aut_Q $ of the quiver  $Q$ consists of sequences of mutations, permutations of the nodes, and inversions, that preserve the quiver $Q$.

As already mentioned, affine Weyl translations (\ref{eq:affine-translations}) have a distinguished role through their appearance as time flows in the theory of discrete Painlev\'e equations. 
In general, a translation $T\in \CT(\widehat{\mathfrak{g}})$ acts on simple roots $\alpha_i$ as follows \cite{2015arXiv150908186K}
\begin{align}\label{eq:Weyl-translations-general}
T_{\vec n}(\alpha_i)=\alpha_i+n_i\delta, && \delta=\sum_im_i\alpha_i, && \sum n_im_i=0\,,
\end{align}
for some choice of $n_i$. 
For $dP_n$ there are $n$ solutions to the last equation above, corresponding  to the existence of $n$ independent flows
\be
	\rk \CT(E_{n}^{(1)}) = n\,.
\ee
The action of a generic translation can be obtained as a sequence of elementary reflections on the root lattice, possibly followed by a permutation of the roots. See \cite[Section 3.2]{2015arXiv150908186K} for a convenient way to compute the translation in terms of reflections.

Once a translation $T\in \CT(E_{n}^{(1)})$ is realized as an element  of $\Aut_Q$, its action gets extended from  $\Gamma_f$ to the  whole $\Gamma$.
The key data encoding this extended action is a collection of \emph{limiting rays} in $\Gamma$.
A limiting ray is defined as $\delta_j \IZ\subset \Gamma$, where 
\be\label{eq:limiting-ray-def}
	\delta_j = \lim_{k\to\infty} \frac{1}{k} T^k(\gamma_i)
\ee
for some  charge $\gamma_i$ labeling the $i$-th node of the quiver. The collection of all limiting rays is obtained by applying the above definition to all $\gamma_i$ (some $\gamma_i$'s may have the same limiting ray, other maybe left invariant  by translations).

\subsection{Stability condition for local del Pezzos from affine translations}\label{Sec:StabRoots}

We now come to the main point of this section, namely to explain the relevance of (extended) affine Weyl group symmetries arising within the Cremona group of $X$, for the study  of BPS states encoded by a quiver $Q$.
As emphasized in Section \ref{sec:MutFlows}, there is an important distinction between affine Weyl translations $\CT$ and the rest of $\Aut_Q$, which arises naturally in the study of discrete Painlev\'e equations. 
Furthermore these equations and the time evolution described by $\CT(\widehat{\mathfrak{g}})$ also play a role in the study of BPS spectrum of  the quiver $Q$.
In particular, \cite{Bonelli:2020dcp} showed explicitly how affine translations
\footnote{More precisely, the double cover $2\cdot \CT$, see below.} 
$\CT(A_1^{(1)})$ reproduce the sequence of mutations for the BPS quiver of $\IF_0$ discussed in \cite{Closset:2019juk}, and conjectured that this observation should generalize to the other del Pezzo geometries by studying the action of discrete flows on the BPS charges of local $\mathbb{F}_1$ and local $dP_3$.

In this paper we wish to make the above proposal more precise, identifying
a relation  between affine translations and certain  stability conditions. Furthermore, in all examples we considered it turns out that the class of stability conditions arising from affine Weyl translations often belongs to a collimation chamber, as introduced in Section~\ref{sec:symmetry-chambers}. This fact will allow us to compute the wall-crossing invariant order by order as outlined in Section \ref{sec:U-constrained}. 
Our goal is to associate uniquely a collimation chamber to an appropriate Weyl translation
\be\label{eq:T-C-map}
	T \mapsto \CC_{T}\,.
\ee

A priori it is unclear whether a Weyl translation $T$ should correspond to a stability condition.
In fact,  through (\ref{eq:Aut-W}) a translation $T$ may generically be realized as a sequence of mutations, permutations, and possibly certain involutions on $Q$. 
There are three basic conditions that $T$ must  satisfy, in order for (\ref{eq:T-C-map}) to make sense:
\begin{enumerate}[label={(c\arabic*)}]
\item \label{criterion-1}
There is a positive integer $k\in \mathbb{N}$ such that 
$
	T^k
	= \mu_{i_1}\dots \mu_{i_\ell} 
$
admits an expression involving only left mutations.
\item\label{criterion-2}
There exists a configuration of $Z_{\gamma_i}\in \IC$ such that clockwise tilting the positive half-plane induces a sequence of mutations identical to iterations of $(T^{k})^n$ for $n\geq 1$.
\item\label{criterion-3}
The above configuration of central charges belongs to a collimation chamber.
\end{enumerate}
If criteria  \ref{criterion-1}-\ref{criterion-2} are  satisfied, we can associate a stability condition  to $T$.  The third requirement is only needed to  restrict to stability  conditions of collimation  type.

In general it is not always clear how to verify whether an affine translation $T$ (more precisely, its realization in  $\Aut_Q$) satisfies the above criteria. 
\ref{criterion-1} is the easiest to verify: for example if $T$ only involves left mutations and pemutations one takes $k$ to be the order of the permutation $\pi^k=1$. Then one uses  $\mu_i\pi = \pi \mu_{\pi(i)}$ to take all instances of $\pi$ together so as to cancel them, leaving mutations only.
Criteria \ref{criterion-2}-\ref{criterion-3} are more difficult to verify, and we will now outline a procedure tailored to this purpose.

We shall  begin with some technical assumptions:
\begin{enumerate}[label={(a\arabic*})]
\item \label{assumption-1}
In the mapping (\ref{eq:flavor-roots}),
 each  simple root $\alpha_i\in\CQ(\hat{\mathfrak{g}})$ is realized as a linear combination of basic charges  
\be\label{eq:roots-local-charges}
	\alpha=\sum_{i} c_{\alpha,i}\gamma_i
\ee
with nonzero coefficients only for a subset  of mutually local $\gamma_i$.
\item \label{assumption-2}
We require that $\langle\delta_i,\delta_j\rangle = 0$ for all charges of limiting rays of $T$.  
\item \label{assumption-3} 
There is a nontrivial sublattice whose roots are given by only mutually local charges. 
\end{enumerate}
If the first condition is not satisfied, we can still define a virtual stability condition, but it will not necessarily correspond to a collimation chamber. This second assumption is only technical, and we hope to be able to lift it in future work. It certainly does not hold in the four-dimensional case, see e.g. the example of the Kronecker quiver in Section \ref{sec:Coll}. For the case of local del Pezzos, it fails for $dP_1,dP_2,dP_4$. We now explain the procedure that associates a collimation chamber to each affine traslation.

\paragraph{Procedure.} 
Given the above assumptions \ref{assumption-1}-\ref{assumption-3}, we can now state a procedure to verify whether conditions \ref{criterion-2}-\ref{criterion-3} are verified. We shall proceed constructively, \emph{i.e.} by stating a set of constraints on the possible configurations of $Z_{\gamma_i}$ that  would realize  \ref{criterion-2}-\ref{criterion-3}.
Let $\varphi_i=\arg(Z_{\gamma_i})$, then we require

\begin{enumerate}
\item
The D0-brane central charge is real 
\begin{equation}\label{eq:Coll0}
	\arg(Z_\delta)=0.
\end{equation}

\item
The limiting rays of $T$ should lie on the real axis 
\begin{equation}\label{eq:Coll1}
	\arg Z_{\delta_i}=0 \qquad\forall i.
\end{equation}
Thanks  to \ref{assumption-2}, this does not correspond to a wall of marginal stability. 
\item
Since roots $\alpha_i$ are made of mutually local charges (see \ref{assumption-1}), we demand that 
$Z_{\gamma_i}$ of the central charges of $\gamma_i$ contributing to (\ref{eq:roots-local-charges}) with nonzero coefficients are mutually aligned
\begin{equation}\label{eq:Coll2}
\varphi_{\alpha,1}=\dots=\varphi_{\alpha, k}\,.
\end{equation}
\end{enumerate}

In the examples that we consider this procedure constrains the $Z_{\gamma_i}$  to the  extent that it completely  fixes the map (\ref{eq:T-C-map}).

\section{Examples}\label{sec:examples}

\subsection{\texorpdfstring{$\IF_0$}{F0}}

\begin{figure}
\begin{center}
\includegraphics[width=.35\textwidth]{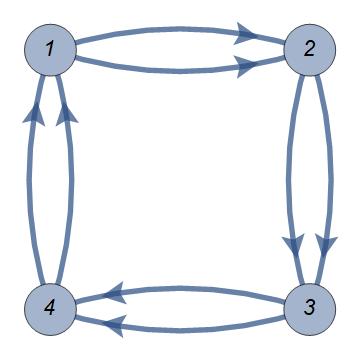}
\caption{BPS quiver for local $\mathbb{F}_0$.
}
\label{Fig:QuiverF0}
\end{center}
\end{figure}

The Cremona group of $\IF_0$ is
\be
	\Cr(\IF_0) = W(A_1^{(1)}) \rtimes {\rm Dih}_4\,.
\ee
The extended affine Weyl group
\be
	 \widetilde{W}(A_1^{(1)}) = W(A_1^{(1)}) \rtimes \Out(A_1^{(1)})
\ee
is a subgroup of $\Cr(\IF_0)$, since $\Out(A_1^{(1)}) \simeq \IZ^2\subset {\rm Dih}_4$ as depicted  in  Figure \ref{Fig:A1Dynkin}.

On the other hand, the full Cremona group is visible on the  BPS quiver, which is depicted in Figure \ref{Fig:QuiverF0}. The adjacency matrix is 
\begin{equation}\label{eq:F0-quiver-matrix}
B=\left( \begin{array}{cccc}
0 & -2 & 0 & 2 \\
2 & 0 & -2 & 0 \\
0 & 2 & 0 & -2 \\
-2 & 0 & 2 & 0
\end{array} \right).
\end{equation}
The full automorphism  group of $Q$ coincides with the Cremona group \cite{Bershtein:2017swf}
\be
	\Aut_Q = \Cr(\IF_0)\,,
\ee
with  generators 
\begin{align}\label{eq:WA1Gens}
s_0=(1,3)\mu_1\mu_3, && s_1=(2,4)\mu_2\mu_4, && \pi=(1,2,3,4), && \sigma=\iota(1,3).
\end{align}
Automorphisms $\Aut_Q$ have a permutation subgroup $\Pi_Q\simeq \IZ_4 \subset {\rm Dih_4}$. 

The rank of $B$ is $2$, therefore
\be
	\rk \Gamma_g=\rk\Gamma_f=2\,.
\ee
A positive integral basis for $\Gamma_f^+ = \{\gamma\in\Gamma_f, {\rm Re}Z_\gamma >0\}$ is given by
\be\label{eq:F0-flavor-generators}
	\gamma_{f,0}=\gamma_1+\gamma_3\,, \qquad
	\gamma_{f,1}=\gamma_2+\gamma_4\,.
\ee

\begin{figure}
\begin{center}
\includegraphics[width=.5\textwidth]{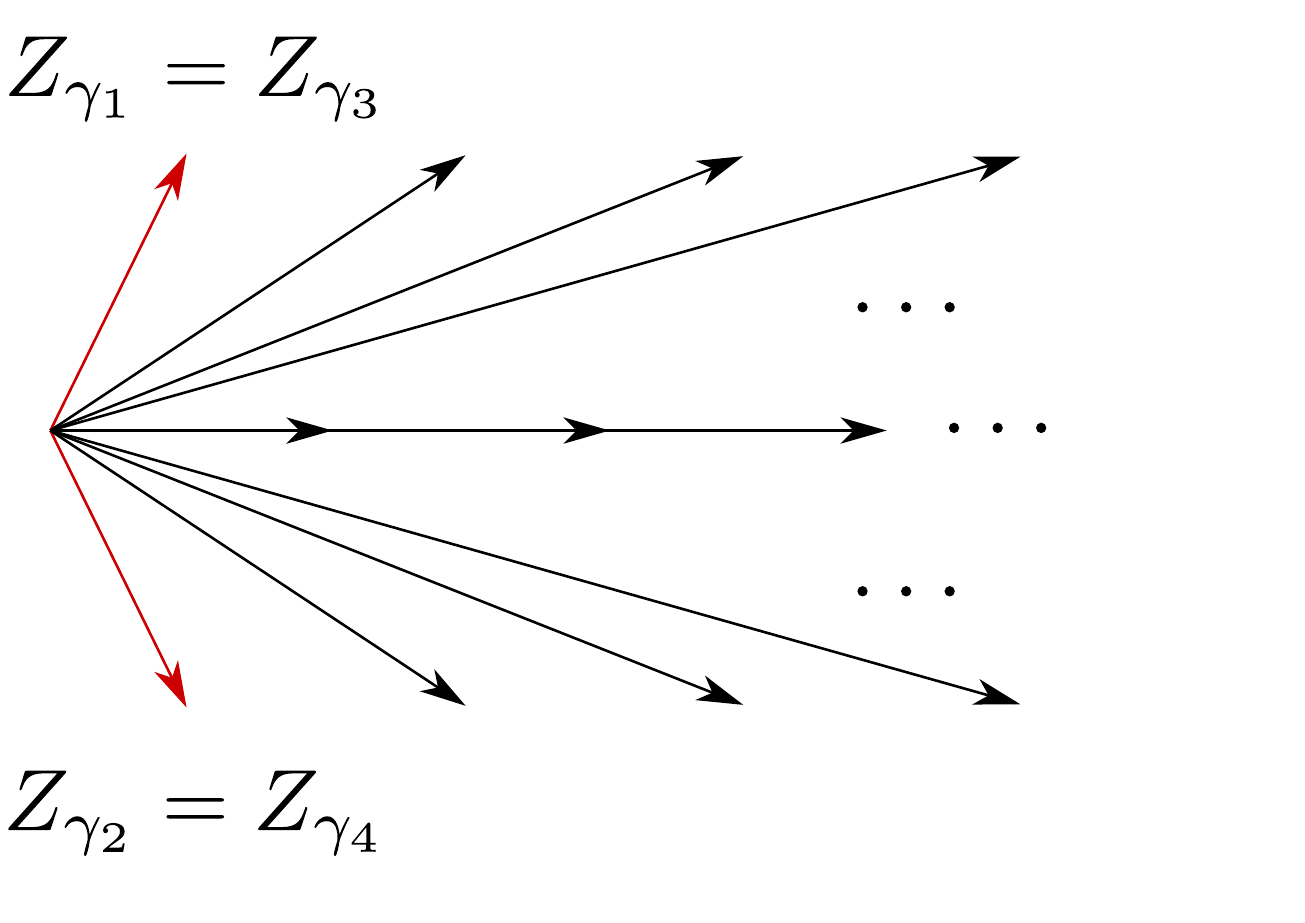}
\caption{BPS ray diagram for the collimation chamber $\CC_1[\IF_0]$.}
\label{fig:F0-raydiagram}
\end{center}
\end{figure}

\begin{figure}
\begin{center}
\begin{tikzpicture}

\draw[fill=SecondBlue,thick] (-1,0) circle (.3);
\draw[fill=SecondBlue,thick] (1,0) circle (.3);
\draw[-,thick] (-0.7,0.1) to (0.7,0.1);
\draw[-,thick] (-0.7,-0.1) to (0.7,-0.1);
\draw[->,thick] (-0.8,0.4) to[out=30,in=150] (0.8,0.4);
\draw[<-,thick] (-0.8,-0.4) to[out=330,in=210] (0.8,-0.4);
\node at (-1,0) {$\alpha_0$};
\node at (1,0) {$\alpha_1$};
\node at (0,1) {$\pi$};
\node at (0,-1) {$\pi$};

\end{tikzpicture}
\end{center}
\caption{$A_1^{(1)}$ Dynkin diagram and automorphisms}
\label{Fig:A1Dynkin}
\end{figure}
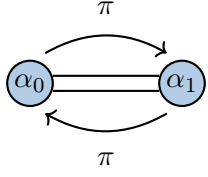

\paragraph{Collimation chambers.}
The moduli space of stability conditions  for the quiver of Figure~\ref{Fig:QuiverF0} features collimation chambers. 
The first one of these, found in \cite{Closset:2019juk}, is a chamber containing the following  configuration  of central  charges\footnote{Recall that by convention all $Z_{\gamma_i}$ in this paper are understood to have positive real part. The full extent of  the chamber depends on its BPS spectrum, in this case $\CC_1$ also contains more general configurations, such as $Z_{\gamma_1}\neq Z_{\gamma_3}$ as long as their phases are the same.
}
\be\label{eq:Closset-DelZotto-chamber}
	\CC_1[\IF_0]\supset
	\quad
	Z_{\gamma_1} = Z_{\gamma_3} \,,
	\quad
	Z_{\gamma_2} = Z_{\gamma_4} \,,
	\quad
	\arg Z_{\gamma_1} > \arg Z_{\gamma_2}\,,
	\quad
	Z_{\gamma_1}+Z_{\gamma_2}\in \IR^+\,.
\ee
The authors of \cite{Closset:2019juk} observed that tilting the positive half-plane of central charges induces infinite sequences of left- and right-mutations, corresponding to the following towers of hypermultiplets 
\be\label{eq:F0-Closset-DelZotto-chamber-hypers}
\begin{split}
	&
	\Omega(\gamma_1+n(\gamma_1+\gamma_2), \CC_1)=1
	\qquad
	\Omega(\gamma_2+n(\gamma_1+\gamma_2), \CC_1)=1
	\\
	&
	\Omega(\gamma_3+n(\gamma_3+\gamma_4), \CC_1)=1
	\qquad
	\Omega(\gamma_4+n(\gamma_3+\gamma_4), \CC_1)=1
\end{split}
\ee
These BPS states are those that lie off the real axis in the ray diagram of Figure \ref{fig:F0-raydiagram}. As $n\to\infty$ rays accumulate towards $Z_{\gamma_1+\gamma_2} = Z_{\gamma_3+\gamma_4} \in \IR^+$ as required by the definition of collimation chamber.

There is also a second collimation chamber, characterized by
\be\label{eq:F0-2nd-chamber}
	\CC_2[\IF_0]\supset
	\quad
	Z_{\gamma_1} = Z_{\gamma_3} \,,
	\quad
	Z_{\gamma_2} = Z_{\gamma_4} \,,
	\quad
	\arg Z_{\gamma_1} < \arg Z_{\gamma_2}\,,
	\quad
	Z_{\gamma_1}+Z_{\gamma_2}\in \IR^+ \,.
\ee
Again there are infinite towers of hypermultiplets accumulating towards $\IR^+$, however their charges are now
\be\label{eq:F0-2nd-chamber-hypers}
\begin{split}
	&
	\Omega(\gamma_1+n(\gamma_1+\gamma_4), \CC_2)=1
	\qquad
	\Omega(\gamma_2+n(\gamma_2+\gamma_3), \CC_2)=1
	\\
	&
	\Omega(\gamma_3+n(\gamma_2+\gamma_3), \CC_2)=1
	\qquad
	\Omega(\gamma_4+n(\gamma_1+\gamma_4), \CC_2)=1 \,.
\end{split}
\ee
In either of these chambers  the spectrum  is not exhausted by hypermultiplets, as there are BPS states  with $Z\in  \IR_+$ that need to be  determined. We'll return to this below.

Clearly, the two chambers $\CC_1, \CC_2$ in (\ref{eq:Closset-DelZotto-chamber}) and (\ref{eq:F0-2nd-chamber}) are related to each other  by  exchanging $\{\gamma_1,\gamma_3\}$ with $\{\gamma_2,\gamma_4\}$ (as unordered sets). 
This is a $\IZ_2$ subgroup of $\Pi_Q$, whose action is therefore not faithful on $\CC_i$.

\paragraph{Translations of the $A_1^{(1)}$ lattice.}
Before proceeding with the analysis of BPS states on the real ray, let us pause to explain how the collimation chambers described above could have been found by studying translations of the affine root lattice associated to the quiver. 
The relation between the chamber (\ref{eq:Closset-DelZotto-chamber}) and lattice translations was first observed by  \cite{Bonelli:2020dcp}. We begin by reviewing how translations of the  root lattice are realized by mutations and  permutations on quiver vertices.

The low-energy gauge theory phase is $5d$ $\CN=1$ pure $SU(2)$ Yang-Mills. The isomorphism  (\ref{eq:flavor-roots}) identifies the flavor sublattice $\Gamma_f$ with the root lattice $\CQ(A_1^{(1)})$.
Simple roots are identified with  the generators (\ref{eq:F0-flavor-generators})
\begin{align}\label{eq:root-flavor-isom-F0}
	\alpha_0=\gamma_{f,0}, && \alpha_1=\gamma_{f,1}.
\end{align}
Therefore the null root $\delta:= \sum_{i=1}^{4}\gamma_i$ decomposes into $\delta =\sum_i m_i \alpha_i $ with $(m_0, m_1)=(1,1)$.
Applying the general definition (\ref{eq:Weyl-translations-general}) of affine Weyl translations we therefore find a single translation generator $T_1$, defined by $(n_0, n_1) = (1,-1)$
acting as follows on the simple roots
\be\label{eq:T1-F0-roots}
	T^n_1(\vec{\alpha})=\left( \begin{array}{c}
                \alpha_0+n\delta \\
                \alpha_1-n\delta
	\end{array} \right)\,.
\ee
Therefore in the case of local $\IF_0$
\be\label{eq:F0-translation-group}
	\CT(A_1^{(1)}) \simeq \IZ\,.
\ee
Next recall the relation (\ref{eq:Aut-W}) between the extended Weyl group of the affine root lattice,  and quiver automorphisms. 
The translation subgroup $\CT(A_1^{(1)})  \subset \widetilde{W}({A_1^{(1)}})$ is realized in $\Aut_Q$ by \cite{Bershtein:2017swf}
\begin{equation}\label{eq:F0Flow1}
	T_1=(1,2)(3,4)\mu_1\mu_3
\end{equation}
Note that while the action of $\CT(A_1^{(1)})$ is only defined on $\CQ(A_1^{(1)}) \simeq \Gamma_f$, the realization (\ref{eq:F0Flow1}) extends to all of $\Gamma$, as  follows
\begin{align}\label{eq:T1-F0-charges}
            T^n_1(\vec{\gamma})=\left( \begin{array}{c}
            \gamma_1+n(\gamma_1+\gamma_2) \\
            \gamma_2-n(\gamma_1+\gamma_2) \\
            \gamma_3+n(\gamma_3+\gamma_4) \\
            \gamma_4-n(\gamma_3+\gamma_4)
            \end{array} \right)\,.
\end{align}
As a check, when restricted to $\gamma_{f,i}$  with $i=0,1$, the action reproduces that on $\alpha_i$ for $i=0,1$.

The affine Weyl group contains only one set of translations (\ref{eq:F0-translation-group}). However its realization  on the BPS charge lattice $\Gamma$ isn't unique. 
This is because the map (\ref{eq:root-flavor-isom-F0}) can be composed with the nontrivial outer automorphism $\pi \in \Out(A_1^{(1)}) \simeq \IZ_2$, which exchanges the two roots 
\be\label{eq:pi-F0}
	\pi:\qquad
	\alpha_0\mapsto\alpha_1\,,
	\ \ 
	\alpha_1\mapsto\alpha_0\,.
\ee
Note that $\Out(A_1^{(1)})\subset \widetilde W(A_1^{(1)})$  belongs to  the extended affine Weyl group, and therefore it must admit a realization as an element  of $\Aut_Q$, by virtue of (\ref{eq:Aut-W}), so that $\pi$ is really the one defined in \eqref{eq:WA1Gens}.
This means that $\Out(A_1^{(1)})$ is realized as a subgroup of quiver permutations
\be\label{eq:out-aut-rel-F0}
	 \mathbb{Z}_2\approx\Out(A_1^{(1)}) \subset \Pi_Q\approx\mathbb{Z}_4 \,,
\ee
since $\pi^4=id$, but $\pi^2|_{\Gamma_f}=id$.

Conjugation by $\pi \in \Out(A_1^{(1)})$ leads to the \emph{same} group of affine translations, but with a \emph{different} realization in $\Aut_Q$. In fact the generator
\begin{equation}\label{eq:F0Flow2}
T_2\equiv(1,4)(2,3)\mu_2\mu_4= \pi^{-1} T_1  \pi.
\end{equation}
acts simply as the  inverse of $T_1$ on $\CQ(A_1^{(1)})$
\be\label{eq:T2-F0-roots}
	T^n_2(\vec{\alpha})=\left( \begin{array}{c}
\alpha_0-n\delta \\
\alpha_1+n\delta
\end{array} \right)\,,
\ee
which  is  the image under $\pi$ of (\ref{eq:T1-F0-roots}).
However for the action on $\Gamma$ we obtain 
\begin{align}\label{eq:T2-F0-charges}
        T^n_2(\vec{\gamma})=\left( \begin{array}{c}
        \gamma_1-n(\gamma_1+\gamma_4) \\
        \gamma_2+n(\gamma_3+\gamma_2) \\
        \gamma_3-n(\gamma_3+\gamma_2) \\
        \gamma_4+n(\gamma_1+\gamma_4)
        \end{array} \right)\,,
\end{align}
which is the image of (\ref{eq:T1-F0-charges}) under $\pi$.

Note that, while $T_2=T_1^{-1}$ on $\CQ(A_1^{(1)})\simeq\Gamma_f$, this is not so on $\Gamma$. This is due  to the fact that  we chose a  realization  of $\Cr(\IF_0)$ in $\Aut_Q$ which only involves \emph{left} mutations, as explained in Section \ref{sec:MutFlows}.

\paragraph{Collimation chambers from affine Weyl translations.} 
Having reviewed the realization of affine Weyl translations, we are now in a position to explain how these give rise to a sequence of mutations which corresponds to a stability condition. Namely, the  sequence of mutations induced by lattice  translations coincides with  the sequence of mutations induced by tilting the positive half-plane of central charges for a certain choice of stability condition.
Furthermore, this stability condition belongs to one  of  the collimation chambers~$\CC_i$. 

The conditions \eqref{eq:Coll0} and \eqref{eq:Coll2} coming from the root lattice structure of $\Gamma$ read in this case
\begin{equation}\label{eq:flow-phases-F0}
\begin{cases}
\varphi_1+\varphi_2+\varphi_3+\varphi_4=0, \\
\varphi_1=\varphi_3, \\
\varphi_2=\varphi_4.
\end{cases}
\end{equation}
To discuss \eqref{eq:Coll1} we must determine the charges $\delta_i$ of limiting rays (\ref{eq:limiting-ray-def}).
For the flow $T_1$, it is clear from (\ref{eq:T1-F0-charges}) that
\be
	\delta_1{(T_1)} = \gamma_1+\gamma_2\,,
	\qquad
	\delta_2{(T_1)} = \gamma_3+\gamma_4\,,
\ee
whereas for the flow $T_2$ in (\ref{eq:T2-F0-charges}) we have instead
\be
	\delta_1{(T_2)} = \gamma_4+\gamma_1\,,
	\qquad
	\delta_2{(T_2)} = \gamma_2+\gamma_3\,.
\ee
In either case we find that $Z_{\delta_i} \in  \IR^+$ thanks to (\ref{eq:flow-phases-F0}). 
Therefore (\ref{eq:Coll1})  is automatically satisfied for both flows. 
The general solution to (\ref{eq:flow-phases-F0}) is
\begin{equation}\label{eq:PhasesF0}
\varphi_1=\varphi_3=-\varphi_2=-\varphi_4.
\end{equation}

\begin{figure}[h]
\begin{center}
\begin{subfigure}{.35\textwidth}
\centering
\includegraphics[width=\textwidth]{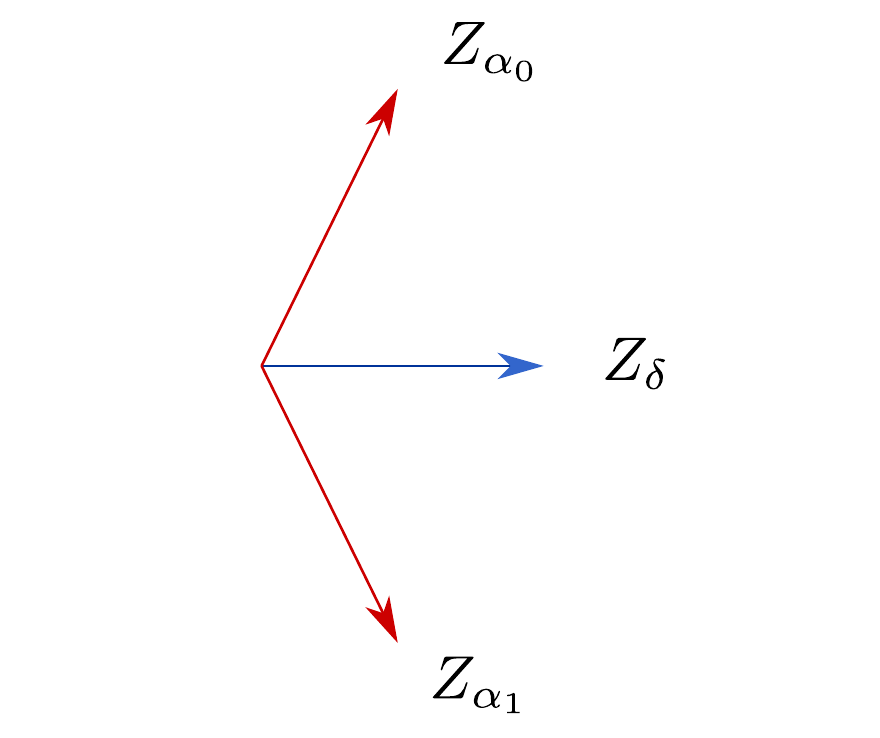}
\caption{Stability condition associated to $T_1$}
\label{Fig:F01}
\end{subfigure}\hspace*{.15\textwidth}
\begin{subfigure}{.35\textwidth}
\centering
\includegraphics[width=\textwidth]{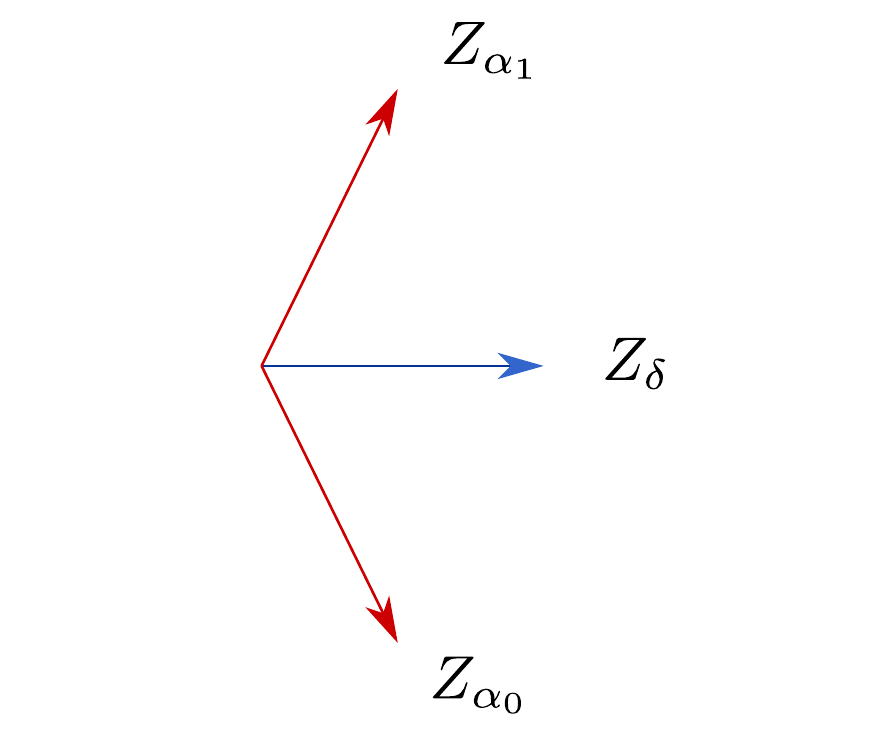}
\caption{Stability condition associated to $T_2$}
\label{Fig:F02}
\end{subfigure}
\caption{Stability conditions for $\IF_0$, corresponding to $\CC_1[\IF_0]$ and $\CC_2[\IF_0]$  respectively.}
\label{Fig:F0Stability}
\end{center}
\end{figure}

However equation \eqref{eq:PhasesF0} does not yet determine completely a stability condition, as it does not fully specify the phase ordering of the central charges. 
This last piece of information is fixed in different ways for the two flows $T_1, T_2$, by viewing each of them as  tiltings of the positive half $Z$-plane. 
Indeed, it was observed in  \cite{Bonelli:2020dcp} that
\begin{align}
	T_1^2 = \mu_2\mu_4\mu_3\mu_1 \equiv  \textbf{m}_1 
	&& 
	T_2^2=\mu_1\mu_3\mu_2\mu_4  \equiv \textbf{m}_2 \,.
\end{align}
For $T_1$ the first mutations occur on $\gamma_1,\gamma_3$, implying
\begin{align}
 \varphi_1=\varphi_3>\varphi_2=\varphi_4=-\varphi_1\,.
\end{align}
The configuration of BPS rays for simple roots is shown in Figure \ref{Fig:F01}. 
This class of stability conditions corresponds exactly to the collimation chamber defined in (\ref{eq:Closset-DelZotto-chamber})
\be
	\text{Stability conditions for }T_1 \equiv \text{collimation chamber }\CC_1[\IF_0]\,.
\ee
Conversely, the first charges mutated by the flow $T_2$ are $\gamma_2,\gamma_4$, so that
\begin{align}
	\varphi_2=\varphi_4>\varphi_1=\varphi_3=-\varphi_2 \,.
\end{align}
The configuration of BPS rays for simple roots is shown in Figure \ref{Fig:F02}. 
This class of stability conditions corresponds exactly to the collimation chamber defined in (\ref{eq:F0-2nd-chamber})
\be
	\text{Stability conditions for }T_2 \equiv \text{collimation chamber }\CC_2[\IF_0]\,.
\ee

\paragraph{Wall-crossing invariant.}
Let us now return  to the analysis of the  two collimation chambers (\ref{eq:Closset-DelZotto-chamber}) and   (\ref{eq:F0-2nd-chamber}),  and to the computation of the wall-crossing  invariant $\IU$.
First we make two observations
\begin{enumerate}
\item\label{obs:F0-1}
The chamber $\CC_2$ in (\ref{eq:F0-2nd-chamber}) and the states (\ref{eq:F0-2nd-chamber-hypers}) are related to $\CC_1$ in (\ref{eq:Closset-DelZotto-chamber}) and its BPS states (\ref{eq:F0-Closset-DelZotto-chamber-hypers}) by a $\Pi_Q\simeq \IZ_4$ shift in charge  labels $\pi:\gamma_i\mapsto\gamma_{i+1\,{\rm mod}4}$. Note that  $\pi^2$ acts trivially on stability conditions, so  the  action of $\Pi_Q\simeq \IZ_4$ is not faithful on $\{\CC_i\}$. 
Indeed, the subgroup $\Out(A_1^{(1)}) \simeq\IZ_2$ is the one that acts faithfully on $\{\CC_1,\CC_2\}$.
\item\label{obs:F0-2} 
Stability conditions (\ref{eq:F0-Closset-DelZotto-chamber-hypers}) and (\ref{eq:F0-2nd-chamber}) are connected by a path in moduli space along which no BPS rays exit the half-plane ${\rm Re}Z>0$. Concretely, let $Z^{(0)}_{\gamma_i}$ be arbitrary representatives for the chamber (\ref{eq:Closset-DelZotto-chamber}). Then a suitable path would be
\be\label{eq:F0-stability-path}
\begin{split}
	Z_{\gamma_1}(t) =  Z_{\gamma_3}(t) = t\cdot Z^{(0)}_{\gamma_1} +(1-t) \cdot Z^{(0)}_{\gamma_2} \\
	Z_{\gamma_2}(t) =  Z_{\gamma_4}(t) = t\cdot Z^{(0)}_{\gamma_2} +(1-t) \cdot Z^{(0)}_{\gamma_1} \\
\end{split}
\qquad t\in[0,1]
\ee
\end{enumerate}
Existence of a path with these properties implies that $\IU$ remains unchanged in moving from  one stability condition to the other. 
Therefore
\be\label{eq:U-Uprime-equality-F0}
	\IU(\CC_1) = \IU(\CC_2)
\ee
where each side admits a  factorization of the form (\ref{eq:collimation-factorization})
\be\label{eq:collimation-factorization-F0}
	\IU(\CC_i) = \IU(\measuredangle^+,\CC_i)\cdot \IU(\IR^+,\CC_i) \cdot \IU(\measuredangle^-,\CC_i)\,.
\ee 
The BPS rays off the real axis contribute, respectively  in each chamber
\be\label{eq:collimation-factors-ClossetDelZotto}
\begin{split}
	\IU(\measuredangle^+,\CC_1)
	& = \prod_{k\geq 0}^{\nearrow}\Phi( X_{\gamma_1 + k (\gamma_1+\gamma_2)}) \Phi( X_{\gamma_3+ k (\gamma_3+\gamma_4)}) 
	\\
	\IU(\measuredangle^-,\CC_1)
	& = \prod_{k\geq 0}^{\searrow}\Phi( X_{\gamma_2 + k (\gamma_1+\gamma_2)}) \Phi( X_{\gamma_4+ k (\gamma_3+\gamma_4)}) 
\end{split}
\ee
and
\be
\begin{split}
	\IU(\measuredangle^+,\CC_2)
	& = \prod_{k\geq 0}^{\nearrow}\Phi( X_{\gamma_2 + k (\gamma_2+\gamma_3)}) \Phi( X_{\gamma_4+ k (\gamma_4+\gamma_1)}) 
	\\
	\IU(\measuredangle^-,\CC_2)
	& = \prod_{k\geq 0}^{\searrow}\Phi( X_{\gamma_3 + k (\gamma_2+\gamma_3)}) \Phi( X_{\gamma_1+ k (\gamma_4+\gamma_1)}) 
\end{split}
\ee
where $\nearrow$ ($\searrow$) denotes increasing (decreasing) values of $k$ to the right.

To compute $\IU(\IR^+,\CC_1)$  and $ \IU(\IR^+,\CC_2)$, consider a formal series expansion like (\ref{U-real-expansion}) for each of them.
Plugging these into (\ref{eq:collimation-factorization-F0}), and imposing the equality  (\ref{eq:U-Uprime-equality-F0}) leads to an equation involving $c_\gamma(\CC_1)$ on the \emph{l.h.s.} and $c_\gamma(\CC_2)$ on the \emph{r.h.s.}
Next we use the fact that factorizations $\IU(\CC_1)$ and $\IU(\CC_2)$ correspond to stability conditions related by a $\IZ_4$ relabeling of $\gamma_i$ on quiver nodes. 
In particular this implies $c_\gamma(\CC_2) = c_{\pi^{-1}(\gamma)}(\CC_1)$ (cf. (\ref{eq:symmetry-derivation})) and we can express both $\IU(\CC_i)$ in terms of $c_\gamma(\CC_1) \equiv c_\gamma$ only.
Wall-crossing invariance (\ref{eq:U-Uprime-equality-F0}) then imposes equations on these coefficients: namely 
\be\label{eq:F0-symmetry-equations}
	\IU(\measuredangle^+,\CC_1) \(  \sum_{\gamma\in \Gamma_{\IR^+}} c_\gamma \, X_\gamma \) \IU(\measuredangle^-,\CC_1)
	=
	\IU(\measuredangle^+,\CC_2) \(  \sum_{\gamma\in \Gamma_{\IR^+}} c_{\pi^{-1}(\gamma)} \, X_\gamma \) \IU(\measuredangle^-,\CC_2) \,.
\ee
This is the equation we needed to constrain $c_\gamma$ and compute $\IU(\IR^+,\CC_i)$.
Expanding each side, this is precisely the symmetry constraint derived in (\ref{eq:symmetry-equation}).
Note that this condition holds for coefficients $u_\gamma$, which are nontrivial functions of the $c_\gamma$ through (\ref{eq:collimation-factorization}) and the known factors (\ref{eq:collimation-factors-ClossetDelZotto}).

To solve equation (\ref{eq:F0-symmetry-equations}) we expand each side as a series  in $X_\gamma$ and compare coefficients of the monomials. This leads to infinitely many equations which can be solved recursively by introducing a filtration in $|\gamma| = \sum_i d_i$.
Such a procedure can be easily implemented on a computer, and yields
\be
\begin{split}
	\IU(\IR^+,\CC_1) 
	& = 
	1
	-\frac{1+y^2}{1-y^2} \( X_{\gamma_1+\gamma_2} +X_{\gamma_3+\gamma_4}\)
	+\frac{2 y^2 + y^4+y^6}{(1-y^2)(1-y^4)} \(X_{2\gamma_1+2\gamma_2} +X_{2\gamma_3+2\gamma_4} \)
	\\
	&
	-\frac{\left(y^{4}+2 y^6+y^{10}\right)}{(1-y^2)^2 (1-y^6)}
	\(X_{3\gamma_1+3\gamma_2} +X_{3\gamma_3+3\gamma_4} \)
	\\
	&
	-\frac{\left(1-y^2\right) \left(1-y^4\right) c_{\gamma_{D0}}-2y^2-3 y^4 + y^6 }{\left(1-y^2\right)^3}
	\(X_{\gamma_1+\gamma_2+\gamma_{D0}}  + X_{\gamma_3+\gamma_4+\gamma_{D0}}\)
	\\
	&
	+c_{\gamma_{D0}} X_{\gamma_{D0}} + \dots
\end{split}	
\ee
where we introduced the shorthand notation $\gamma_{D0} = \sum_i\gamma_i$.

The solution can be obtained to arbitrary orders in $|\gamma|$, and we find that it is  compatible with the factorization
\be
	\IU_0(\IR^+,\CC_1) 
	=
	\prod_{s=\pm1}\prod_{k\geq 0}\Phi((-y)^{s} X_{\gamma_1+\gamma_2+ k \gamma_{D0}})^{-1}
	\cdot \Phi((-y)^{s} X_{\gamma_3+\gamma_4+ k \gamma_{D0}})^{-1}
\ee
where $\gamma_{D0}  =  \sum_{i=1}^{4}\gamma_i$.
We can write this in a more suggestive way by  recalling the D-brane charges associated with $\gamma_i$ as follows
\be
	\IU_0(\IR^+,\CC_1) 
	=
	\prod_{s=\pm1}\prod_{k\geq 0}\Phi((-y)^{s} X_{\gamma_1+\gamma_2+ k \gamma_{D0}})^{-1}
	\cdot \Phi((-y)^{s} X_{ - \gamma_1-\gamma_2 +(k+1) \gamma_{D0}})^{-1}\,.
\ee
Indeed recall that $\gamma_1+\gamma_2$ can be identified with a D2 brane wrapping the fiber $\IP^1$ in $\IF_0$, then the  above expression makes it evident that these are towers of $D2\-D0$ boundstates as well as the CPT conjugate  $\overline{D2}\-D0$ states which belong to the positive half $Z$-plane \cite{Longhi:2021qvz}. We will see this pattern arising in greater generality in other examples below.

This corresponds to a BPS spectrum along the real ray consisting of vectormultiplets
\be\label{eq:F0-real-ray-states}
	\Omega(\gamma_1+\gamma_2+n\gamma_{D0},\CC_1) = \Omega(-\gamma_1-\gamma_2+(n+1)\gamma_{D0},\CC_1) = y+y^{-1}
\ee
As observed in \cite{Longhi:2021qvz}, this formula misses contributions from D0 branes, an ambiguity reflected in the  fact that the coefficients $c_{n\gamma_{D0}}$ remain undetermined.
Since $\gamma_{D0}$ has trivial pairing with any  other charge $\langle\gamma_{D0},\gamma\rangle=0$, it does not participate in wall-crossing. It follows that pure D0 states cannot be fixed by our methods, which  are  fundamentally based on  symmetries of $Q$ in conjunction with wall-crossing  invariance.
The  D0 contributions can be obtained with other techniques, such as exponential networks \cite{Eager:2016yxd, Banerjee:2018syt, Banerjee:2019apt, Banerjee:2020moh, Longhi:2021qvz} or a combination of results from Attractor Flow Tree formulae and known results about DT invariants \cite{Alexandrov:2018iao, Beaujard:2020sgs, Mozgovoy:2020has, Mozgovoy:2021iwz}. 
The expected motivic DT invariant for any number of D0 branes is \cite{Mozgovoy:2020has}
\be
	\Omega(n\gamma_{D0};y) = y^3 + 2y+y^{-1}\,.
\ee
The corresponding contribution to $\IU(\IR^+)$ is a product of quantum dilogarithms as outlined in (\ref{eq:factorization}) \emph{et seq.}
\be
	\IU(\IR^+,\CC_1)  = \IU_0(\IR^+,\CC_1) \cdot 
	\prod_{n\geq 1} \Phi((-y)^{-1} X_{n\gamma_{D0}})^{-1} \Phi((-y)^{} X_{n\gamma_{D0}})^{-2} \Phi((-y)^{3} X_{n\gamma_{D0}})^{-1}\,.
\ee
It is worth recalling a subtle distinction in the counting of D0 branes: the Donaldson-Thomas counting is based on compactly-supported de Rham cohomolgy of quiver representation varieties, while the physical counting of BPS states  (pertaining e.g. to gauge theory in the context of geometric engineering) is based on the $L^2$ cohomology \cite{Lee:1997vp, Yi:1997eg}. The latter typically contains fewer states, and leads to Poincar\'e-symmetric expressions for  $\Omega(\gamma,y)$ while the former does not, see \cite{Duan:2020qjy, Mozgovoy:2020has} for a discussion.

In conclusion, the spectrum we obtain for local $\mathbb{F}_0$ in chamber  $\CC_1$ is
\begin{equation}\label{eq:SpectrumF0}
\begin{array}{|c|c|}
	\hline
	\gamma & \Omega(\gamma;y) \\
	\hline\hline
	\gamma_1 + k (\gamma_1+\gamma_2) & 1\\
	-\gamma_1 + (k+1) (\gamma_1+\gamma_2) & 1\\
	\gamma_3+ k (\gamma_3+\gamma_4) & 1\\
	-\gamma_3+ (k+1) (\gamma_3+\gamma_4) & 1\\
	\hline
	\gamma_1+\gamma_2+k\gamma_{D0} & y+y^{-1} \\
	-\gamma_1-\gamma_2+(k+1)\gamma_{D0} & y+y^{-1} \\
	(k+1)\gamma_{D0} & y^3 + 2y+y^{-1}\\
	\hline
\end{array}
\end{equation}
with $k\geq  0$, plus the respective antiparticles obtained by sending $\gamma\to-\gamma$.  This manifestly agrees with  the explicit expression in  \cite{Longhi:2021qvz}, it can  also be shown  to agree with the generating function of stacky invariants obtained in \cite{Mozgovoy:2020has}.\footnote{We thank Sergey Mozgovoy and Boris Pioline for correspondence on this check.}

It should be noted that the spectrum is organized as \emph{two} copies of the spectrum of 4d pure $SU(2)$ SYM from the subquivers shown in Figure \ref{fig:F0-subquivers}, in which the local $\mathbb{F}_0$ quiver is divided by the time flows,  with additional towers of D0 branes over the vector multiplet states.
\begin{figure}[h]
\begin{center}
\begin{subfigure}{.35\textwidth}
\centering
\includegraphics[width=\textwidth]{figures/F0Sub.jpg}
\caption{Four-dimensional subquivers for the flow $T_1$ of $\mathbb{F}_0$.}
\label{Fig:dP51}
\end{subfigure}\hspace*{.15\textwidth}
\begin{subfigure}{.35\textwidth}
\centering
\includegraphics[width=\textwidth]{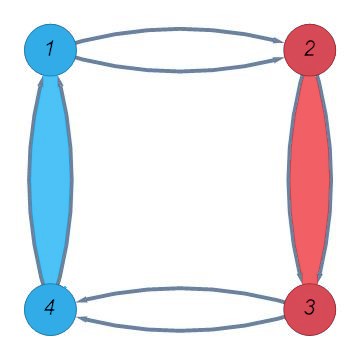}
\caption{Four-dimensional subquivers for the flow $T_2$ of $\mathbb{F}_0$.}
\label{Fig:dP52}
\end{subfigure}
\caption{}
\label{fig:F0-subquivers}
\end{center}
\end{figure}

\begin{remark}
Equation (\ref{eq:U-Uprime-equality-F0}) translates into the following wall-crossing identity
\be
\begin{split}
	& \prod_{k\geq 0}^{\nearrow}\Phi( X_{\gamma_1 + k (\gamma_1+\gamma_2)}) \Phi( X_{\gamma_3+ k (\gamma_3+\gamma_4)}) 
	\\
	 &\qquad  \times
	\prod_{s=\pm1}\prod_{k\geq 0}\Phi((-y)^{s} X_{\gamma_1+\gamma_2+ k \gamma_{D0}})^{-1}
	\cdot \Phi((-y)^{s} X_{\gamma_3+\gamma_4+ k \gamma_{D0}})^{-1}
	\\
	 & \qquad  \qquad \times \prod_{k\geq 0}^{\searrow}\Phi( X_{\gamma_2 + k (\gamma_1+\gamma_2)}) \Phi( X_{\gamma_4+ k (\gamma_3+\gamma_4)}) 
	\\
	= &  \prod_{k\geq 0}^{\nearrow}\Phi( X_{\gamma_2 + k (\gamma_2+\gamma_3)}) \Phi( X_{\gamma_4+ k (\gamma_4+\gamma_1)}) 
	\\
	&\qquad \times 
	\prod_{s=\pm1}\prod_{k\geq 0}\Phi((-y)^{s} X_{\gamma_4+\gamma_1+ k \gamma_{D0}})^{-1}
	\cdot \Phi((-y)^{s} X_{\gamma_2+\gamma_3+ k \gamma_{D0}})^{-1}
	\\
	&\qquad \qquad\times \prod_{k\geq 0}^{\searrow}\Phi( X_{\gamma_3 + k (\gamma_2+\gamma_3)}) \Phi( X_{\gamma_1+ k (\gamma_4+\gamma_1)}) 
	\\
\end{split}
\ee
where we factored out the contribution of  pure-flavor states (pure D0 branes), which  do not participate in wall-crossing. We have tested this as an identity in the numerical ($y\to -1$) setting, by acting  on $X_{\gamma_i}$ and expanding as formal series in $X_\gamma$ with $|\gamma|\leq 12$.
\end{remark}

\subsection{\texorpdfstring{$dP_3$}{dP3}}
\begin{figure}[h]
\begin{center}
\includegraphics[width=.35\textwidth]{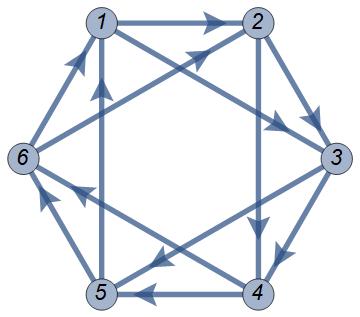}
\end{center}
\caption{BPS quiver for local $\text{dP}_3$}
\label{Fig:QuiverdP3}
\end{figure}
The BPS quiver for the theory corresponding to local $dP_3$ is depicted in Figure \ref{Fig:QuiverdP3}, and has adjacency matrix
\begin{equation}
B=
\left(
\begin{array}{cccccc}
 0 & -1 & -1 & 0 & 1 & 1 \\
 1 & 0 & -1 & -1 & 0 & 1 \\
 1 & 1 & 0 & -1 & -1 & 0 \\
 0 & 1 & 1 & 0 & -1 & -1 \\
 -1 & 0 & 1 & 1 & 0 & -1 \\
 -1 & -1 & 0 & 1 & 1 & 0 \\
\end{array}
\right)
\end{equation}
The permutation symmetry of this  quiver is $\Pi_Q\simeq \IZ_6$. The adjacency matrix has rank 2, hence $\rk \Gamma_f=4$.

\paragraph{Translations of the $(A_1+ A_2)^{(1)}$ lattice.}
The low-energy gauge theory phase of this theory is $SU(2)$ Super Yang-Mills with two fundamental hypermultiplets, and the affine root lattice  is $\Gamma_f\simeq \CQ((A_1+ A_2)^{(1)})$.
We can take as its generators the following vectors in $\Gamma$: 
\begin{align}\label{eq:dP3-alpha-roots}
\alpha_0=\gamma_3+\gamma_6, && \alpha_1=\gamma_1+\gamma_4, && \alpha_2=\gamma_2+\gamma_5,
\end{align}
\begin{align}\label{eq:dP3-beta-roots}
\beta_0=\gamma_2+\gamma_4+\gamma_6, && \beta_1=\gamma_1+\gamma_3+\gamma_5.
\end{align}
The null root is again $\delta = \sum_i\gamma_i$.

Note how the affine root lattice $\CQ((A_1+ A_2)^{(1)})$ has rank four, with generators the simple roots of $A_2$, the simple root of $A_1$ and the null root $\delta$ which is `shared'. This is in line with the statement that such an affine root lattice is isomorphic to $\Gamma_f$ which also has rank four. As a matter of fact, this is reflected in the explicit form of the isomorphism (\ref{eq:flavor-roots}) given above, where it is clear that the five generators obey the relation
\be
	\alpha_0+\alpha_1+\alpha_2 = \beta_0+\beta_1 \equiv \delta\,.
\ee 

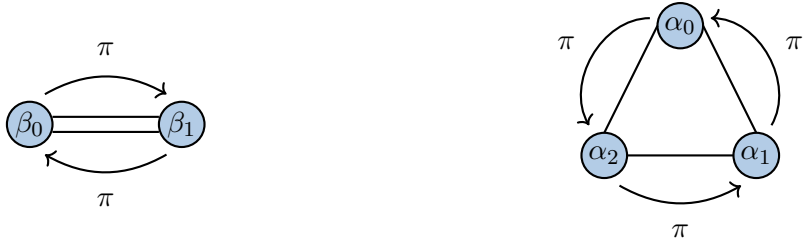
\begin{figure}
\begin{center}
\begin{subfigure}{.5\textwidth}
\centering
\begin{tikzpicture}

\draw[fill=SecondBlue,thick] (-1,0) circle (.3);
\draw[fill=SecondBlue,thick] (1,0) circle (.3);
\draw[-,thick] (-0.7,0.1) to (0.7,0.1);
\draw[-,thick] (-0.7,-0.1) to (0.7,-0.1);
\draw[->,thick] (-0.8,0.4) to[out=30,in=150] (0.8,0.4);
\draw[<-,thick] (-0.8,-0.4) to[out=330,in=210] (0.8,-0.4);
\node at (-1,0) {$\beta_0$};
\node at (1,0) {$\beta_1$};
\node at (0,1) {$\pi$};
\node at (0,-1) {$\pi$};

\end{tikzpicture}
\end{subfigure}\hfill
\begin{subfigure}{.5\textwidth}
\centering
\begin{tikzpicture}

\draw[fill=SecondBlue,thick] (-1,0) circle (.3);
\draw[fill=SecondBlue,thick] (1,0) circle (.3);
\draw[fill=SecondBlue,thick] (0,1.72) circle (.3);
\draw[-,thick] (-0.7,0) to (0.7,0);
\draw[-,thick] (-1,0.3) to (-0.3,1.72);
\draw[-,thick] (1,0.3) to (0.3,1.72);
\draw[<-,thick] (-1.2,0.4) to[out=30+90,in=180] (-0.4,1.82);
\draw[->,thick] (-0.8,-0.4) to[out=330,in=210] (0.8,-0.4);
\draw[->,thick] (1.2,0.4) to[out=-30+90,in=0] (0.4,1.82);
\node at (-1,0) {$\alpha_2$};
\node at (1,0) {$\alpha_1$};
\node at (0,1.72) {$\alpha_0$};
\node at (-1.5,1.5) {$\pi$};
\node at (1.5,1.5) {$\pi$};
\node at (0,-1) {$\pi$};

\end{tikzpicture}
\end{subfigure}
\end{center}
\caption{$(A_2+A_1)^{(1)}$ Dynkin diagrams and automorphisms. $\pi^2=id$ on the $A_1^{(1)}$ sublattice, while $\pi^3=id$ on the $A_2^{(1)}$ sublattice. The two combine to give the $\mathbb{Z}_6$ permutation symmetry of the quiver.}
\label{Fig:A1Dynkin2}
\end{figure}

In this case the Cremona group coincides with the extended Weyl group 
\be
	\Cr(dP_3)\simeq \widetilde W((A_1+ A_2)^{(1)}) = W((A_1+ A_2)^{(1)}) \rtimes {\rm Dih}_6 \,.
\ee
This is generated by the reflections on the $A_2,A_1$ sublattices
\begin{align}\label{eq:dP3A2Refl}
s_0=(3,6)\mu_6\mu_3 && s_1=(1,4)\mu_4\mu_1, && s_2=(2,5)\mu_5\mu_2,
\end{align}
\begin{align}
r_0=(4,6)\mu_2\mu_4\mu_6\mu_2, && r_1=(3,5)\mu_1\mu_3\mu_5\mu_1,
\end{align}
and by the outer automorphisms generating ${\rm Dih}_6$
\begin{align}
\pi=(1,2,3,4,5,6), && \sigma=(1,4)(2,3)(5,6)\iota\,.
\end{align}
Of these, $\pi$ is an order-six Dynkin diagram automorphism that permutes simple roots, see Figure \ref{Fig:A1Dynkin2}. This is the generator of $\Pi_Q\subset {\rm Dih}_6$ that we'll use to constrain $\IU$.

We will focus here only on the $A_2^{(1)}$ sublattice spanned by $\alpha_1,\alpha_2,\alpha_3$, since it is the only one whose roots contain only locally commuting charges, so that it gives rise to a collimation chamber as discussed in Section \ref{Sec:StabRoots}. (The sublattice $A_1^{(1)}$ does not satisfy our technical assumption \ref{assumption-1}, we leave its discussion to future work.)
The subgroup of affine Weyl translations 
\be
	\IZ^3\simeq \CT(A_2^{(1)}) \subset \CT((A_1+ A_2)^{(1)} )
\ee
is realized in $\Aut_Q$ by the following generators
\begin{align} \label{eq:dP3A2Flows}
	T_1=\pi^2s_2s_1, && T_2=\pi^2s_0s_2, && T_3=\pi^2s_1s_0\,.
\end{align}
Its action on the sublattice of the affine root lattice $\Gamma_f$ corresponding to $\CQ(A_2^{(1)})$ 
is 
\begin{align}
        T_1^n(\vec{\alpha})=\left( \begin{array}{c}
        \alpha_0-n\delta, \\
        \alpha_1+n\delta, \\
        \alpha_2
        \end{array} \right), &&
        T_2^n(\vec{\alpha})=\left( \begin{array}{c}
        \alpha_0, \\
        \alpha_1-n\delta, \\
        \alpha_2+n\delta
        \end{array} \right) &&
        T_3^n(\vec{\alpha})=\left( \begin{array}{c}
        \alpha_0+n\delta, \\
        \alpha_1, \\
        \alpha_2-n\delta
        \end{array} \right).
        \end{align}
This action extends to  $\Gamma$ as follows
\be
\begin{split}\label{eq:T1T2-dP3-charges}
        & T_{1}^n(\vec{\gamma})=
        \left(\begin{array}{c}
        \gamma_1+n(\gamma_1+\gamma_2+\gamma_3),\\
        \gamma_2, \\ 
        \gamma_3-n(\gamma_1+\gamma_2+\gamma_3),\\
        \gamma_4+n(\gamma_4+\gamma_5+\gamma_6), \\
        \gamma_5,\\
        \gamma_6-n(\gamma_4+\gamma_5+\gamma_6)
        \end{array}\right)\,, 
        \qquad  
        T_2^n(\vec{\gamma})=
        \left(\begin{array}{c}
        \gamma_1-n(\gamma_1+\gamma_5+\gamma_6) \\
        \gamma_2+ n(\gamma_2+\gamma_3+\gamma_4) \\
        \gamma_3 \\
        \gamma_4 -n(\gamma_2+\gamma_3+\gamma_4) \\
        \gamma_5+n(\gamma_1+\gamma_5+\gamma_6) \\
        \gamma_6.
        \end{array}\right)\,,
        \\
        &
        \qquad\qquad\qquad\qquad
        T_3^n(\vec{\gamma})=
        \left(\begin{array}{c}
        \gamma_1\\
        \gamma_2- n(\gamma_1+\gamma_2+\gamma_6) \\
        \gamma_3 +n(\gamma_3+\gamma_4+\gamma_5)\\
        \gamma_4 \\
        \gamma_5-n(\gamma_3+\gamma_4+\gamma_5)\\
        \gamma_6+n(\gamma_1+\gamma_2+\gamma_6).
        \end{array}\right)\,.
\end{split}
\ee
As in the  case of local $\IF_0$, we observe that the three flows are related by the action of $\IZ_3\subset \Out(A_2^{(1)})$ permuting the three   simple roots $\alpha_i$.%
\footnote{Note that $\pi$ is a $\IZ_6\simeq \Pi_Q$ generator on the full charge lattice $\Gamma$,  and also on the affine root lattice $\CQ((A_1+ A_2)^{(1)})$. 
This $\IZ_6$ is contained in the outer automorphism  group $\Out((A_1+ A_2)^{(1)}) \simeq {\rm Dih}_6$.
Then $\pi^2$ generates a $\IZ_3$ outer automorphisms from $\Out(A_2^{(1)}) \simeq  {\rm Dih}_3$, which acts trivially  on $\CQ(A_1^{(1)})$. %
} 
\begin{equation}
\begin{split}\label{eq:TimesPermutations}
T_1
& = \pi^{-2}T_2\pi^{2}=\pi^{-4}T_3\pi^{4} \, .
\end{split}
\end{equation}

\begin{figure}[h!]
\begin{center}
\begin{subfigure}{.32\textwidth}
\centering
\includegraphics[width=\textwidth]{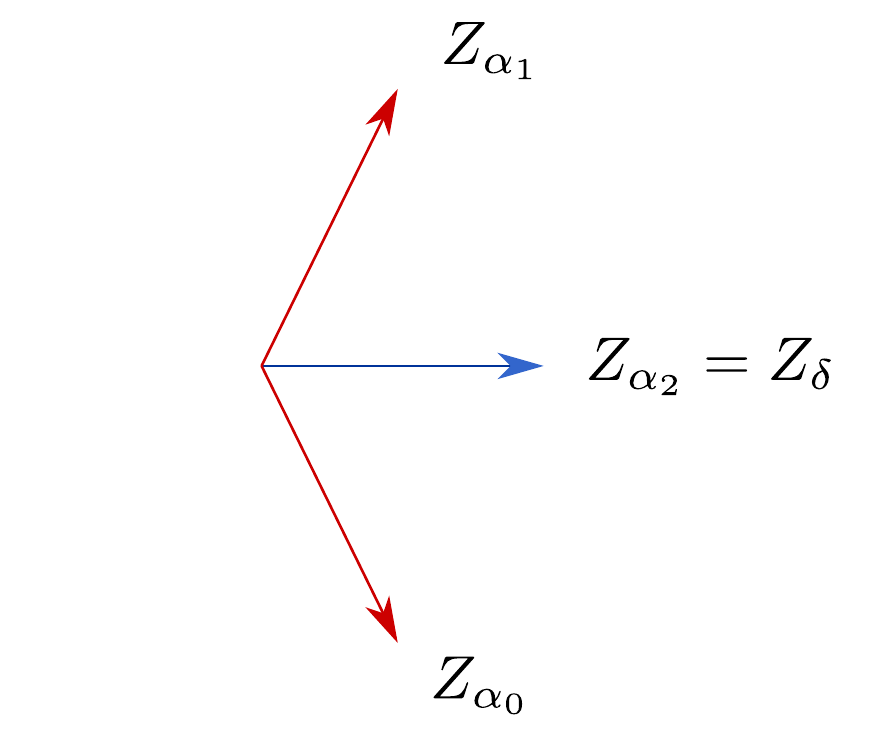}
\caption{Stability condition for $T_1$}
\label{Fig:dP31}
\end{subfigure}\hfill
\begin{subfigure}{.32\textwidth}
\centering
\includegraphics[width=\textwidth]{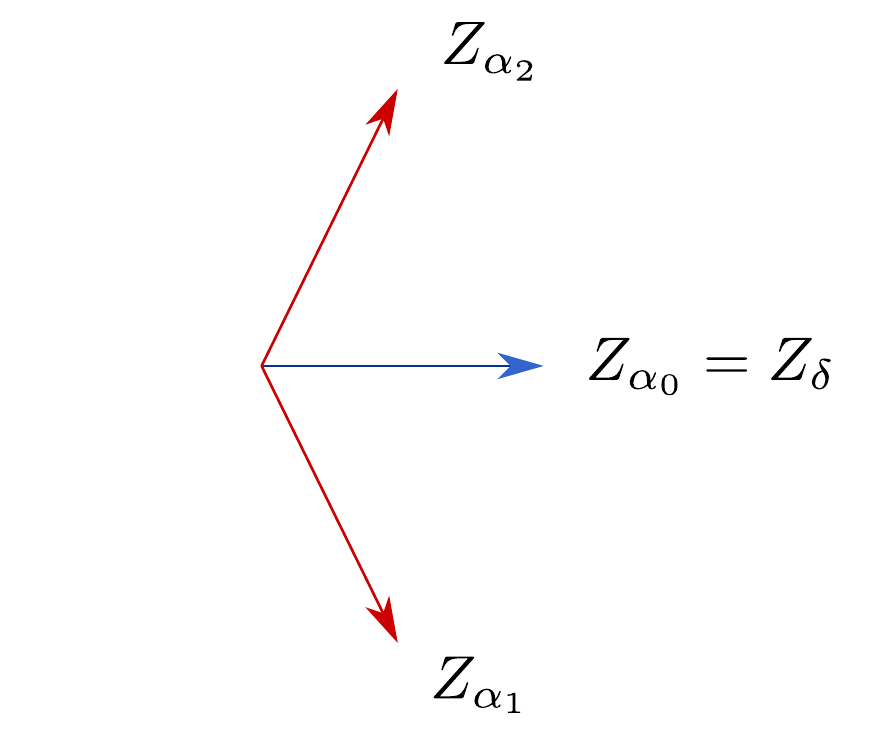}
\caption{Stability condition for $T_2$}
\label{Fig:dP32}
\end{subfigure}\hfill
\begin{subfigure}{.32\textwidth}
\centering
\includegraphics[width=\textwidth]{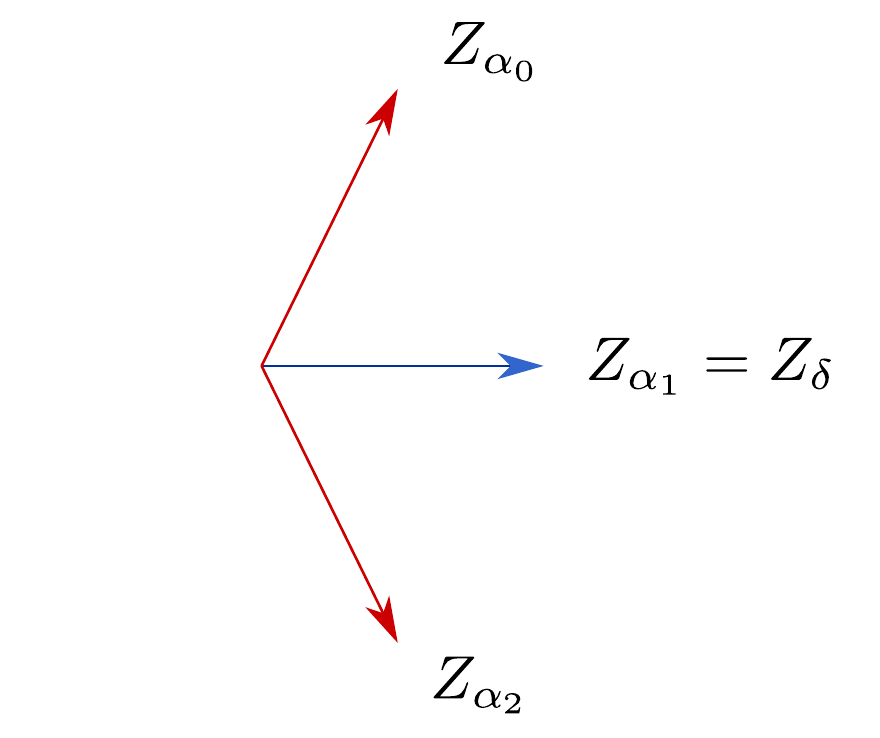}
\caption{Stability condition for $T_3$}
\label{Fig:dP33}
\end{subfigure}
\caption{Stability conditions for $dP_3$.}
\label{Fig:dP3}
\end{center}
\end{figure}

\paragraph{Collimation chambers.} 
Having identified the relevant flows (\ref{eq:dP3A2Flows}), we now study the associated stability conditions following the procedure outlined  in Section \ref{Sec:StabRoots}.
Recall that $\delta = \sum_{i=1}^{6}\gamma_i$, while $\alpha_i$ are expressed in terms of $\gamma_i$ as in (\ref{eq:dP3-alpha-roots}). Then the conditions  corresponding to equations \eqref{eq:Coll0} and \eqref{eq:Coll2} translate here into
\begin{equation}
\begin{cases}
\varphi_1+\varphi_2+\varphi_3+\varphi_4+\varphi_5+\varphi_6=0 \\
\varphi_3=\varphi_6, \\
\varphi_1=\varphi_4, \\
\varphi_2=\varphi_5. \\
\end{cases}
\end{equation}
This system by itself does not yet determine the stability condition. 
Recall  indeed that we impose an additional requirement \eqref{eq:Coll1},  corresponding to demanding that limiting rays lie on the real axis.  The accumulation  rays for the flow $T_1$ in (\ref{eq:limiting-ray-def}) can be read off from~(\ref{eq:T1T2-dP3-charges}) 
\be\label{eq:dP3-T1-limiting-rays}
	\delta_1(T_1) = \gamma_1+\gamma_2+\gamma_3\,,
	\qquad
	\delta_2(T_1) = \gamma_4+\gamma_5+\gamma_6\,,
\ee
therefore we demand
\begin{equation}\label{eq:dP3-T1-limiting-rays-reality}
\varphi_1+\varphi_2+\varphi_3=\varphi_4+\varphi_5+\varphi_6=0.
\end{equation}
As $\gamma_2,\gamma_5$ are invariant under $T_1$, we can think of them in a certain sense as limiting vectors as well. It is then quite natural to also impose
\begin{equation}
\varphi_2=\varphi_5=0.
\end{equation}
Finally, observing that the first charge mutated by $T_1$ in \eqref{eq:dP3A2Flows} is $\gamma_1$, the stability condition corresponding to the first flow is
\begin{align}\label{eq:dP3-stability-conditions}
\text{Stability condition for } T_1: && \varphi_1=\varphi_4>\varphi_2=\varphi_5=0>\varphi_3=\varphi_6=-\varphi_1,
\end{align}
as in Figure \ref{Fig:dP31}.

\begin{remark}
The condition $\theta_2=\theta_5=0$ is not strictly necessary, but we need it to fix everything since $A_2^{(1)}$ is only a sublattice of the full root system. It is suggested by the fact that these states are left invariant by the flow $T_1$ (and so they are strictly speaking limiting vectors for the flow, although they are hypermultiplets). It is consistent since
\be
\begin{split}
	\langle\gamma_2,\gamma_5\rangle&=\langle\gamma_2,\gamma_1+\gamma_2+\gamma_3\rangle
	=\langle\gamma_5,\gamma_1+\gamma_2+\gamma_3\rangle\\
	&=\langle\gamma_2,\gamma_4+\gamma_5+\gamma_6\rangle=\langle\gamma_5,\gamma_4+\gamma_5+\gamma_6\rangle=0.
\end{split}
\ee
\end{remark}
The conditions for the flows $T_2$ and $T_3$ respectively are obtained by applying $\pi^2$ and $\pi^4$ to the charges, so we have the other two stability conditions
\begin{align}\label{eq:dP3-stability-conditions-T2}
\text{Stability condition for } T_2: && \varphi_3=\varphi_6>\varphi_4=\varphi_1=0>\varphi_2=\varphi_5=-\varphi_3,
\end{align}
\begin{align}\label{eq:dP3-stability-conditions-T3}
\text{Stability condition for } T_3: && \varphi_2=\varphi_5>\varphi_3=\varphi_6=0>\varphi_1=\varphi_4=-\varphi_2,
\end{align}
illustrated in Figure \ref{Fig:dP32} and \ref{Fig:dP33} respectively. 

So far we applied the criteria of Section \ref{Sec:StabRoots}  to  deduce the  stability  conditions associated to flows $T_i$. As a  check, we observe that iterations of  the translation  generators indeed admit a realization that only  involves left-mutations
Note that 
\begin{align}
T_i^2=\textbf{m}_i^3,
\end{align}
where
\begin{align}
\textbf{m}_1=\mu_6\mu_3\mu_5\mu_2\mu_4\mu_1, && \textbf{m}_2=\pi^2 \textbf{m}_1\pi^{-2}, && \textbf{m}_3=\pi^4\textbf{m}_1\pi^{-4},
\end{align}
Moreover we claim that iterations of these mutations  correspond to tiltings of the positive half-plane for choices of stability data as given  in  (\ref{eq:dP3-stability-conditions}), (\ref{eq:dP3-stability-conditions-T2}) and (\ref{eq:dP3-stability-conditions-T2}).
Once again, the time flows are permuted version of charge tiltings, and contain the full information about the stability condition.

\paragraph{BPS spectrum and Wall-crossing invariant.}
A configuration of central charges 
satisfying conditions (\ref{eq:dP3-stability-conditions})
corresponding to $T_1$ can be taken as follows
\be
\begin{array}{cc}
\CC_1[dP_3]\supset\qquad
&
\begin{split}	
	& Z_{\gamma_1} = Z_{\gamma_4} \,,
	\quad
	Z_{\gamma_2} = Z_{\gamma_5} \,,
	\quad
	Z_{\gamma_3} = Z_{\gamma_6} \,,
	\\
	& \arg Z_{\gamma_1} > \arg Z_{\gamma_2}> \arg Z_{\gamma_3}\,,
	\quad
	Z_{\gamma_1}+Z_{\gamma_3} = Z_{\gamma_2 }\in \IR^+ \,.
\end{split}
\end{array}
\ee
Similarly we denote by $\CC_i[dP_3]$ the chambers corresponding  to flows $T_i$ with $i=2,3$.
Since the limiting rays of  $T_1$ are (\ref{eq:dP3-T1-limiting-rays}), and since we demanded that  their central charges are  real (\ref{eq:dP3-T1-limiting-rays-reality}),
 the flow is expected to produce towers of states asymptotic to the real axis (also see \cite{Bonelli:2020dcp} for a discussion).
This matches the definition of collimation chamber. 

Having identified the choice of stability condition corresponding to a (putative) collimation chamber, we may deduce parts of the spectrum contained in the sectors $\measuredangle^\pm$ of the $Z$-plane (corresponding to ${\rm Re}Z>0$ and  ${\rm Im}Z>0$ or $<0$) by tilting the  positive half-plane and following the induced sequences of left- and right-mutations. 
The mutation sequences are, respectively
\footnote{These coincide with the mutation sequences observed in \cite[eq. (2.36)]{Bonelli:2020dcp}.  To see it, it suffices to separate all permutations (taking them to the left) from all mutations (taken to the right) in their formulae.}
\be
\begin{split}
	\text{CW tilt} & \quad  \textbf{m}_1 = \mu_6  \mu_3  \mu_5  \mu_2  \mu_4  \mu_1
	\\
	\text{CCW tilt} & \quad  \widetilde{\textbf{m}}_1 = \widetilde{\mu}_4  \widetilde{\mu}_1  \widetilde{\mu}_5  \widetilde{\mu}_2  \widetilde{\mu}_6  \widetilde{\mu}_3
\end{split}	
\ee
Acting on $Q$ with $\textbf{m}_1^n$ and $\widetilde{\textbf{m}}_1^n$ yields respectively the following towers of hypermultiplets
\be
\begin{split}
	\Omega(\gamma_1+n(\gamma_1+\gamma_2+\gamma_3))& =1
	\\
	\Omega(\gamma_3+n(\gamma_1+\gamma_2+\gamma_3))& =1
	\\
	\Omega(-\gamma_1+(n+1)(\gamma_1+\gamma_2+\gamma_3))&=1
	\\
	\Omega(-\gamma_3+(n+1)(\gamma_1+\gamma_2+\gamma_3))&=1
\end{split}
\ee
for $n\geq 0$, and similarly with $\{1,2,3\}\to \{4,5,6\}$.
As promised, their central  charges accumulate towards $\IR^+$ from both sides for $n\to+\infty$.
It is  then straightforward to write down $\IU(\measuredangle^\pm)$ explicitly as a product of phase-ordered quantum dilogarithms.
\be
\begin{split}
	\IU(\measuredangle^+,\CC_1) 
	& = 
	\prod^{\nearrow}_{k\geq 0}
	\Bigg(
	\Phi\(X_{\gamma_1+ k (\gamma_1+\gamma_2+\gamma_3)}\) 
	\Phi\(X_{\gamma_4+ k (\gamma_4+\gamma_5+\gamma_6)}\)
	\\
	&\qquad\qquad
	\times
	\Phi\(X_{\gamma_1+\gamma_2+ k (\gamma_1+\gamma_2+\gamma_3)}\) 
	\Phi\(X_{\gamma_4+\gamma_5+ k (\gamma_4+\gamma_5+\gamma_6)}\)
	\bigg)
	\\ 
	\IU(\measuredangle^-,\CC_1) 
	& = 
	\prod^{\searrow}_{k\geq 0}
	\Bigg(
	\Phi\(X_{\gamma_5+\gamma_6+ k (\gamma_4+\gamma_5+\gamma_6)}\)
	\Phi\(X_{\gamma_2+\gamma_3+ k (\gamma_1+\gamma_2+\gamma_3)}\) 
	\\
	&\qquad\qquad
	\times
	\Phi\(X_{\gamma_6+ k (\gamma_4+\gamma_5+\gamma_6)}\)
	\Phi\(X_{\gamma_3+ k (\gamma_1+\gamma_2+\gamma_3)}\) 
	\Bigg)
\end{split}
\ee

To determine $\IU(\IR^+)$ we adopt a formal series expansion as in (\ref{U-real-expansion}) and 
express $\IU$ as a function of the coefficients $c_\gamma$ of $\IU(\IR^+)$ through (\ref{eq:collimation-factorization}). 
We then impose (\ref{eq:symmetry-equation}) using the $\Pi_Q\simeq \IZ_6$ symmetry of the quiver of Figure \ref{Fig:QuiverdP3} 
\be
	\pi: \, \gamma_{i} \mapsto \gamma_{i+1\, {\rm mod} \, 6}.
\ee 
Solving for the coefficients $c_\gamma$ we obtain a series in $X_\gamma$ up to arbitrary order $|\gamma|$, which is compatible with the  following factorization
\be
\begin{split}
	\IU_0(\IR^+,\CC_1) 
	= \prod_{k\geq 0}\Bigg(
	& 
	\Phi(X_{\gamma_2+ k \gamma_{D0}})\cdot \Phi(X_{\gamma_1+\gamma_3+ k \gamma_{D0}})
	\\
	\cdot 
	&\Phi(X_{-\gamma_2 + (k+1) \gamma_{D0}})\cdot \Phi(X_{-\gamma_1-\gamma_3+ (k+1) \gamma_{D0}})
	\\
	\cdot 
	&\prod_{s=\pm1} \Phi((-y)^{s} X_{\gamma_1+\gamma_2+\gamma_3+ k \gamma_{D0}})^{-1}
	\Bigg)
	\\
	&
	\times \(\{1,2,3\}\to\{4,5,6\}\)
\end{split}	
\ee
where $\gamma_{D0} = \sum_i\gamma_i$.
This corresponds to a spectrum along the real ray consisting of
\be
\begin{split}
	&\Omega(\gamma_2+n\gamma_{D0}) = \Omega(-\gamma_2+(n+1)\gamma_{D0})  = 1\\
	&\Omega(\gamma_5+n\gamma_{D0}) = \Omega(-\gamma_5+(n+1)\gamma_{D0})  = 1\\
	&\Omega(\gamma_1+\gamma_3+n\gamma_{D0}) = \Omega(-\gamma_1-\gamma_3+(n+1)\gamma_{D0}) = 1\\
	&\Omega(\gamma_4+\gamma_6+n\gamma_{D0}) = \Omega(-\gamma_4-\gamma_6+(n+1)\gamma_{D0})  = 1\\
	&\Omega(\gamma_1+\gamma_2+\gamma_3+n\gamma_{D0})  = \Omega(-\gamma_1-\gamma_2-\gamma_3+(n+1)\gamma_{D0})  = y+y^{-1}
\end{split}
\ee
As in the case of local $\IF_0$, the spectrum of BPS states on the real ray features towers of BPS states
corresponding to Kaluza-Klein towers of charges $\gamma_2,\gamma_5,\gamma_1+\gamma_3,\gamma_{4}+\gamma_6$, and their CPT conjugate KK towers.
From the viewpoint of the splitting into two 4d $SU(2)$  $N_f=1$ subquivers, these KK  towers correspond to bound states of $N_f=1$ quarks with D0 branes.

\begin{figure}[b]
\begin{center}
\begin{subfigure}{.25\textwidth}
\centering
\includegraphics[width=\textwidth]{figures/dP3Sub.jpg}
\caption{Four-dimensional subquivers for the flow $T_1$ of $dP_3$.}
\label{Fig:dP51}
\end{subfigure}\hfill
\begin{subfigure}{.25\textwidth}
\centering
\includegraphics[width=\textwidth]{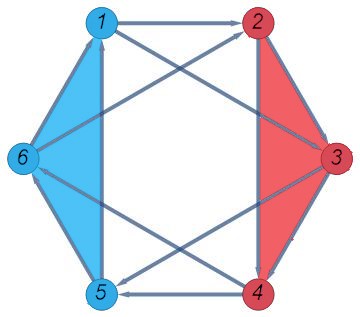}
\caption{Four-dimensional subquivers for the flow $T_2$ of $dP_3$.}
\label{Fig:dP52}
\end{subfigure}\hfill
\begin{subfigure}{.25\textwidth}
\centering
\includegraphics[width=\textwidth]{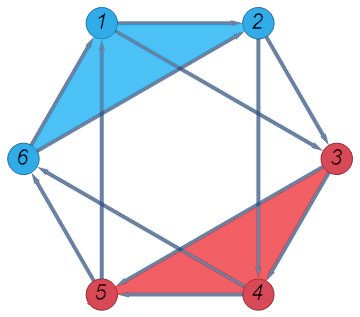}
\caption{Four-dimensional subquivers for the flow $T_3$ of $dP_3$.}
\label{Fig:dP52}
\end{subfigure}
\caption{}
\label{fig:dP3-subquivers}
\end{center}
\end{figure}

Once again this misses the contributions from pure D0 branes. We refer to the  remarks below (\ref{eq:F0-real-ray-states}) for a discussion.
In this case the expected motivic DT invariant for any number of D0 branes is \cite{Mozgovoy:2020has}
\be
	\Omega(n\gamma_{D0};y) = y^3 + 4y+y^{-1}\,.
\ee
The corresponding contribution to $\IU(\IR^+)$ is a product of quantum dilogarithms as outlined in (\ref{eq:factorization}) \emph{et seq.}
\be
	\IU(\IR^+,\CC_1)  = \IU_0(\IR^+,\CC_1) \cdot 
	\prod_{n\geq 1} \Phi((-y)^{-1} X_{n\gamma_{D0}})^{-1} \Phi((-y)^{} X_{n\gamma_{D0}})^{-4} \Phi((-y)^{3} X_{n\gamma_{D0}})^{-1}\,.
\ee

In conclusion, the spectrum we obtain for local $dP_3$ in chamber  $\CC_1$ is
\begin{equation}\label{eq:SpectrumdP3}
\begin{array}{|c|c|}
	\hline
	\gamma & \Omega(\gamma;y) \\
	\hline\hline
	\gamma_r + k (\gamma_1+\gamma_2+\gamma_3) & 1\\
	-\gamma_r + (k+1) (\gamma_1+\gamma_2+\gamma_3) & 1\\
	\gamma_s + k (\gamma_4+\gamma_5+\gamma_6) & 1\\
	-\gamma_s + (k+1) (\gamma_4+\gamma_5+\gamma_6) & 1\\
	\hline
	\gamma_t +k\gamma_{D0} & 1 \\
	-\gamma_t +(k+1)\gamma_{D0} & 1 \\
	\gamma_a+\gamma_b +k\gamma_{D0} & 1 \\
	-\gamma_a-\gamma_b +(k+1)\gamma_{D0} & 1 \\
	\gamma_1+\gamma_2 + \gamma_3+k\gamma_{D0} & y+y^{-1} \\
	-\gamma_1-\gamma_2-\gamma_3+(k+1)\gamma_{D0} & y+y^{-1} \\
	(k+1)\gamma_{D0} & y^3 + 4y+y^{-1}\\
	\hline
\end{array}
\end{equation}
with $k\geq  0$ and
\be
	r\in\{1,3\}
	\qquad 
	s\in\{4,6\}
	\qquad 
	t\in\{2,5\}
	\qquad
	(a,b) \in  \{(1,3), (4,6)\}\,.
\ee
The spectrum  also  includes the respective antiparticles obtained by sending $\gamma\to-\gamma$.

\begin{remark}
The wall-crossing constraint that we solved is the following identity
\be
\begin{split}
	& \prod^{\nearrow}_{k\geq 0}
	\Bigg(
	\Phi\(X_{\gamma_1+ k (\gamma_1+\gamma_2+\gamma_3)}\) 
	\Phi\(X_{\gamma_4+ k (\gamma_4+\gamma_5+\gamma_6)}\)
	\\
	&\qquad\qquad
	\times
	\Phi\(X_{- \gamma_3 + (k+1) (\gamma_1+\gamma_2+\gamma_3)}\) 
	\Phi\(X_{-\gamma_6 + (k+1) (\gamma_4+\gamma_5+\gamma_6)}\)
	\bigg)
	\\
	& \times
	\prod_{k\geq 0}\Bigg(
	\Phi(X_{\gamma_2+ k \gamma_{D0}}) 
	\Phi(X_{-\gamma_2+ (k+1) \gamma_{D0}})
	\Phi(X_{\gamma_1+\gamma_3+ k \gamma_{D0}})
	\Phi(X_{-\gamma_1-\gamma_3+ (k+1) \gamma_{D0}})
	\\ 
	&
	\qquad\times 
	\Phi(X_{\gamma_5+ k \gamma_{D0}}) 
	\Phi(X_{-\gamma_5+ (k+1) \gamma_{D0}})
	\Phi(X_{\gamma_4+\gamma_6+ k \gamma_{D0}})
	\Phi(X_{-\gamma_4-\gamma_6+ (k+1) \gamma_{D0}})
	\\ 
	&\qquad  \times 
	\(\prod_{s=\pm1} 
	\Phi((-y)^{s} X_{\gamma_1+\gamma_2+\gamma_3+ k \gamma_{D0}})^{-1}
	\Phi((-y)^{s} X_{-\gamma_1-\gamma_2-\gamma_3+ (k+1) \gamma_{D0}})^{-1}
	\)
	\Bigg)
	\\
	& \times
	\prod^{\searrow}_{k\geq 0}
	\Bigg(
	\Phi\(X_{-\gamma_4+ (k+1) (\gamma_4+\gamma_5+\gamma_6)}\)
	\Phi\(X_{-\gamma_1+ (k+1) (\gamma_1+\gamma_2+\gamma_3)}\) 
	\\
	&\qquad\qquad
	\times
	\Phi\(X_{\gamma_6+ k (\gamma_4+\gamma_5+\gamma_6)}\)
	\Phi\(X_{\gamma_3+ k (\gamma_1+\gamma_2+\gamma_3)}\) 
	\Bigg)
	\\
	= & 
	\(\{\gamma_1,\gamma_2,\gamma_3,\gamma_4,\gamma_5,\gamma_6\}\to\{\gamma_2,\gamma_3,\gamma_4,\gamma_5,\gamma_6,\gamma_1\}\)
\end{split}\label{eq:WCIDP3}
\ee
where we factored out the contribution of  pure-flavor states (pure D0 branes), which  do not participate in wall-crossing. We have tested this as an identity in the numerical ($y\to -1$) setting, by acting  on $X_{\gamma_i}$ and expanding as formal series in $X_\gamma$ with $|\gamma|\leq 12$.
\end{remark}

\subsection{\texorpdfstring{$dP_5$}{dP5}}\label{Sec:dP5}

\begin{figure}[h]
\begin{center}
\includegraphics[width=.35\textwidth]{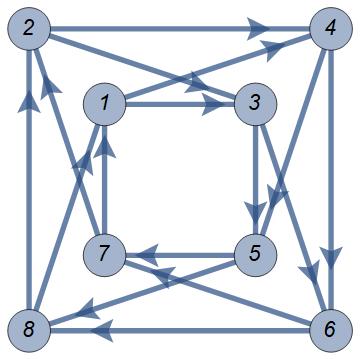}
\end{center}
\caption{BPS quiver for $\text{dP}_5$}
\label{Fig:QuiverdP5}
\end{figure}
The BPS quiver for the theory corresponding to local $dP_5$ is depicted in Figure \ref{Fig:QuiverdP5}, and has adjacency matrix\footnote{This quiver  is related by mutations to the one obtained in \cite[Figure 19]{Hanany:2001py}. 
It  also matches with the quiver in \cite[Figure 8A]{Wijnholt:2002qz}. 
As pointed out in \cite{Wijnholt:2002qz} this is also equivalent to the quiver in Figure 7 of the same paper, which further coincides with the recent derivation in \cite[eq. (8.44)]{Closset:2021lhd}.
Although local $dP_5$ is not toric, this quiver can also be obtained by techniques for toric CY threefolds, see model 4a in \cite{Hanany:2012hi}, which corresponds to a \emph{pseudo} del Pezzo surface $PdP_5$. The distinction between quivers of $dP_5$ and $PdP_5$ lies in the possible  presence of bidirectional arrows for the latter \cite{Beaujard:2020sgs}, however this depends on the superpotential, which plays no  role in our construction.}
\begin{equation}
B=
\left(
\begin{array}{cccccccc}
 0 & 0 & -1 & -1 & 0 & 0 & 1 & 1 \\
 0 & 0 & -1 & -1 & 0 & 0 & 1 & 1 \\
 1 & 1 & 0 & 0 & -1 & -1 & 0 & 0 \\
 1 & 1 & 0 & 0 & -1 & -1 & 0 & 0 \\
 0 & 0 & 1 & 1 & 0 & 0 & -1 & -1 \\
 0 & 0 & 1 & 1 & 0 & 0 & -1 & -1 \\
 -1 & -1 & 0 & 0 & 1 & 1 & 0 & 0 \\
 -1 & -1 & 0 & 0 & 1 & 1 & 0 & 0 \\
\end{array}
\right)
\end{equation}
The permutation symmetry of this  quiver is $\Pi_Q\simeq (\IZ_2)^{\times 4} \rtimes \IZ_4$. 
The adjacency matrix of this quiver has rank two, so $\rk \Gamma_f = 6$.

\paragraph{Translations of the $D_5^{(1)}$ lattice.}
The low-energy gauge theory phase of this theory is $SU(2)$ Super Yang-Mills with four fundamental hypermultiplets, and the affine root lattice is $\CQ(D_5^{(1)})\simeq\Gamma_f$. We can take as its generators the following vectors in $\Gamma$:
\begin{align}\label{eq:D5Roots}
\alpha_0=\gamma_2-\gamma_1, && \alpha_1=\gamma_6-\gamma_5, && \alpha_2=\gamma_1+\gamma_5, \\
\alpha_3=\gamma_3+\gamma_7, && \alpha_4=\gamma_4-\gamma_3, && \alpha_5=\gamma_8-\gamma_7.
\end{align}
The null root is again $\delta = \sum_i\gamma_i$. 

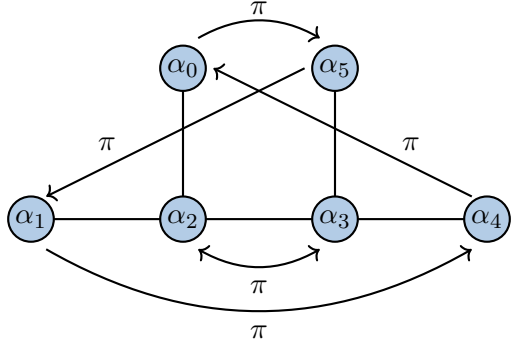
\begin{figure}
\begin{center}
\begin{tikzpicture}

\draw[fill=SecondBlue,thick] (-1,0) circle (.3);
\draw[fill=SecondBlue,thick] (1,0) circle (.3);
\draw[fill=SecondBlue,thick] (-3,0) circle (.3);
\draw[fill=SecondBlue,thick] (3,0) circle (.3);
\draw[fill=SecondBlue,thick] (-1,2) circle (.3);
\draw[fill=SecondBlue,thick] (1,2) circle (.3);
\draw[-,thick] (-0.7,0) to (0.7,0);
\draw[-,thick] (-2.7,0) to (-1.3,0);
\draw[-,thick] (2.7,0) to (1.3,0);
\draw[-,thick] (-1,0.3) to (-1,1.7);
\draw[-,thick] (1,0.3) to (1,1.7);
\draw[->,thick] (-0.8,2.4) to[out=30,in=150] (0.8,2.4);
\draw[<->,thick] (-0.8,-0.4) to[out=330,in=210] (0.8,-0.4);
\draw[->,thick] (-2.8,-0.4) to[out=330,in=210] (2.8,-0.4);
\draw[->,thick] (0.6,2) to (-2.8,0.3);
\draw[<-,thick] (-0.6,2) to (2.8,0.3);
\node at (-3,0) {$\alpha_1$};
\node at (-1,0) {$\alpha_2$};
\node at (1,0) {$\alpha_3$};
\node at (3,0) {$\alpha_4$};
\node at (-1,2) {$\alpha_0$};
\node at (1,2) {$\alpha_5$};
\node at (0,2.8) {$\pi$};
\node at (0,-0.9) {$\pi$};
\node at (-2,1) {$\pi$};
\node at (0,-1.5) {$\pi$};
\node at (2,1) {$\pi$};

\end{tikzpicture}
\end{center}
\caption{$D_5^{(1)}$ Dynkin diagram and automorphisms}
\label{Fig:D5Dynkin}
\end{figure}

The Cremona group coincides with the extended affine Weyl group 
\be
	\Cr(dP_5) = \widetilde{W}(D_5^{(1)}) = {W}(D_5^{(1)}) \rtimes \Out(D_5^{(1)})\,.
\ee
Note that $\Out(D_5^{(1)})$ contains the permutation symmetries $\Pi_Q$ as a subgroup.
Indeed, $\Cr(dP_5)$ can be realized in $\Aut_Q$ as follows
\begin{align}
s_0=(1,2), && s_1=(5,6), && s_2=(1,5)\mu_1\mu_5, \\
s_3=(3,7)\mu_3\mu_7, && s_4=(3,4), && s_5=(7,8),
\end{align}
\begin{align}
\pi=
(1,3,5,7)(2,4,6,8), && \sigma=(1,7)(2,8)(3,5)(4,6)\iota,
\end{align}
where $s_i$ is an elementary reflection along the root $\alpha_i$, and
\begin{equation}
\pi(\vec{\alpha})=\left( \begin{array}{c}
\alpha_4 \\
\alpha_5 \\
\alpha_3 \\
\alpha_2 \\
\alpha_1 \\
\alpha_0
\end{array} \right)
\end{equation}
is the Dynkin diagram automorphism of Figure \ref{Fig:D5Dynkin}. 
All affine  simple roots $\alpha_i$ correspond to sums of mutually local charges, in fulfillment of assumption \ref{assumption-3} in Section \ref{Sec:StabRoots}.

Instead of discussing the whole subgroup of affine Weyl translations $\CT(D_5^{(1)})$, we just focus on the $\IZ$-subgroup generated by 
\begin{align}\label{eq:dP5Transl1}
T_1^n(\vec{\alpha})=\left( \begin{array}{c}
\alpha_0\\
\alpha_1\\
\alpha_2+n\delta\\
\alpha_3-n\delta\\
\alpha_4\\
\alpha_5 
\end{array}
\right).
\end{align}
This admits the following realization in $\Aut_Q$ 
\be
	T_1=s_3s_4s_5s_3s_2s_1s_0s_2(2,6)(1,5)(4,8)(7,3)\,,
\ee
which extends the definition of $T_1$ to the whole charge lattice $\Gamma$ 
\begin{align}\label{eq:dP5Transl1-charges}
T_1^n(\vec{\gamma})=\left( \begin{array}{c}
\gamma_1+n(\gamma_1+\gamma_2+\gamma_3+\gamma_4) \\
\gamma_2+n(\gamma_1+\gamma_2+\gamma_3+\gamma_4) \\
\gamma_3-n(\gamma_1+\gamma_2+\gamma_3+\gamma_4) \\
\gamma_4-n(\gamma_1+\gamma_2+\gamma_3+\gamma_4) \\
\gamma_5+n(\gamma_5+\gamma_6+\gamma_7+\gamma_8) \\
\gamma_6+n(\gamma_5+\gamma_6+\gamma_7+\gamma_8) \\
\gamma_7-n(\gamma_5+\gamma_6+\gamma_7+\gamma_8) \\
\gamma_8-n(\gamma_5+\gamma_6+\gamma_7+\gamma_8) 
\end{array} \right).
\end{align}

We can obtain another, independent flow, by applying the Dynkin diagram automorphism 
\be
	T_2=\pi T_1\pi^{-1}
\ee
acting on roots, and on the whole $\Gamma$, respectively as follows
\begin{align}\label{eq:dP5Transl2}
T_2^n(\vec{\alpha})=\left( \begin{array}{c}
\alpha_0\\
\alpha_1\\
\alpha_2-n\delta\\
\alpha_3+n\delta\\
\alpha_4\\
\alpha_5 
\end{array}
\right),
&&
T_2^n(\vec{\gamma})=\left( \begin{array}{c}
\gamma_1-n(\gamma_1+\gamma_2+\gamma_7+\gamma_8) \\
\gamma_2-n(\gamma_1+\gamma_2+\gamma_7+\gamma_8) \\
\gamma_3+n(\gamma_3+\gamma_4+\gamma_5+\gamma_6) \\
\gamma_4+n(\gamma_3+\gamma_4+\gamma_5+\gamma_6) \\
\gamma_5-n(\gamma_3+\gamma_4+\gamma_5+\gamma_6) \\
\gamma_6-n(\gamma_3+\gamma_4+\gamma_5+\gamma_6) \\
\gamma_7+n(\gamma_1+\gamma_2+\gamma_7+\gamma_8) \\
\gamma_8+n(\gamma_1+\gamma_2+\gamma_7+\gamma_8) 
\end{array} \right).
\end{align}

\begin{remark}
Note that the Weyl translation naturally splits the BPS quiver into two four-dimensional subquivers \cite{Bonelli:2020dcp}.
The towers of states that we find are those appearing in the weakly-coupled spectrum  of these subquivers,  shown in Figure \ref{fig:4d-subquivers}
\end{remark}

\begin{figure}[h]
\begin{center}
\begin{subfigure}{.35\textwidth}
\centering
\includegraphics[width=\textwidth]{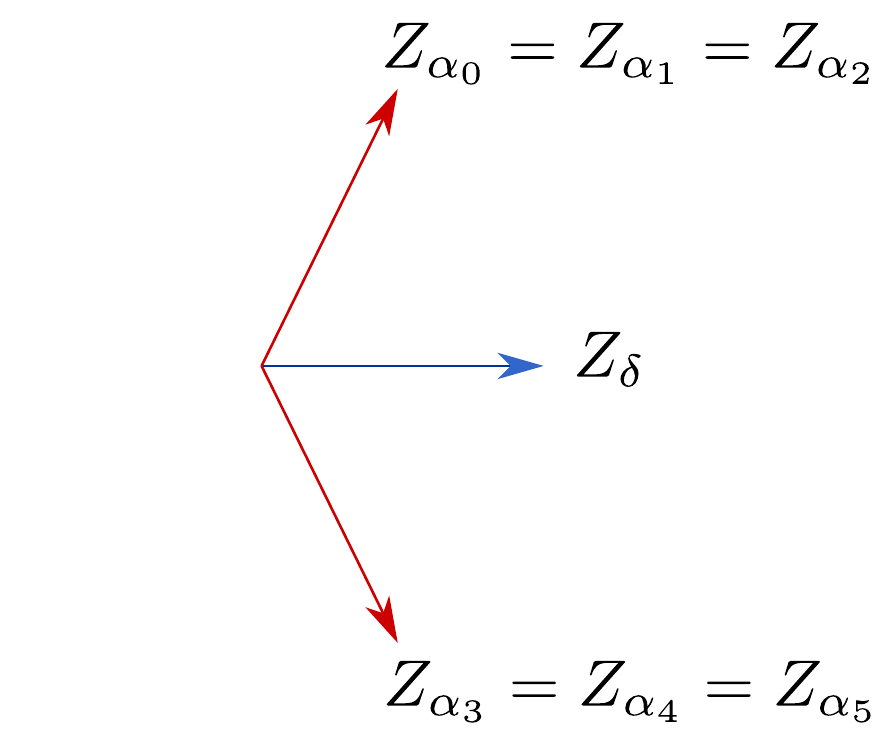}
\caption{Stability condition associated to $T_1$}
\label{Fig:dP51}
\end{subfigure}\hspace*{.15\textwidth}
\begin{subfigure}{.35\textwidth}
\centering
\includegraphics[width=\textwidth]{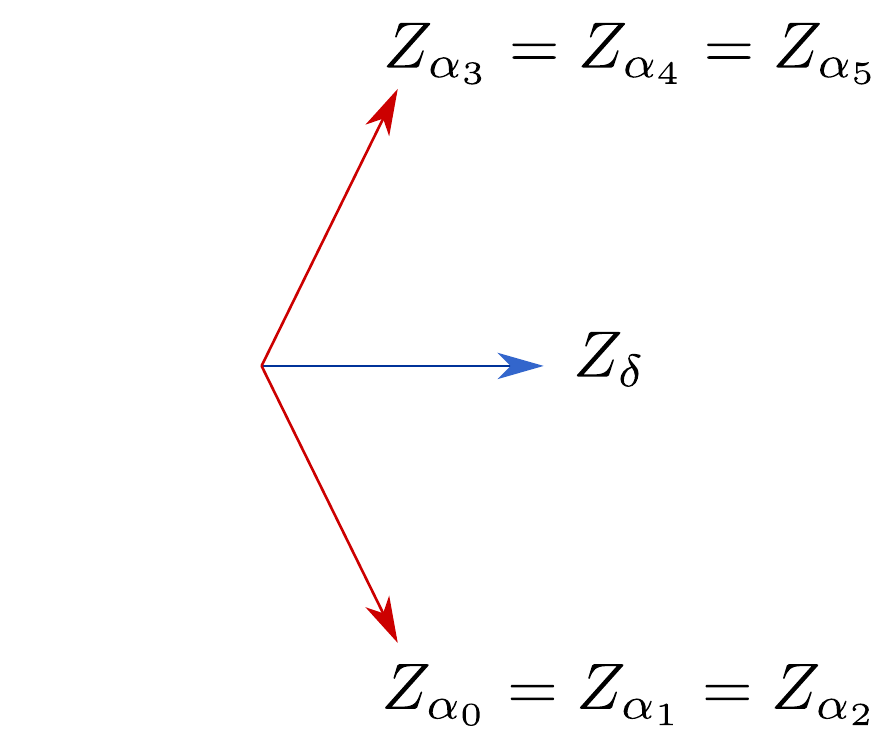}
\caption{Stability condition associated to $T_2$}
\label{Fig:dP52}
\end{subfigure}
\caption{Stability conditions for $dP_5$,  corresponding to  $\CC_1[dP_5]$ and $\CC_2[dP_5]$. 
}
\label{Fig:dP5Stability} 
\end{center}
\end{figure}

\paragraph{Collimation chambers.} 
We now turn to the discussion of collimation chambers associated to the flows $T_1$ and $T_2$.
Recalling that $\delta = \sum_{i=1}^{8}\gamma_i$, while $\alpha_i$ are expressed in terms of $\gamma_i$ as in (\ref{eq:D5Roots}), conditions  corresponding to equations \eqref{eq:Coll0} and \eqref{eq:Coll2} translate here into%
\footnote{{This explains why we considered just the flow $T_1$ instead of the whole translation group $\CT(D_5^{(1)})$. In principle the root lattice $\CQ(D_5^{(1)})$ allows for many different flows. However, the technical condition \eqref{eq:Coll2}, that requires all charges entering in a root in \eqref{eq:D5Roots} to lie on the same ray, greatly restricts our allowed choices of tiltings. It is plausible that by relaxing this condition one could make sense also of the other affine translations.}}
\begin{equation}
\begin{cases}
\sum_i\varphi_i=0,\\
\varphi_2=\varphi_1, \\
\varphi_6=\varphi_5, \\
\varphi_1=\varphi_5, \\
\varphi_3=\varphi_7, \\
\varphi_4=\varphi_3, \\
\varphi_8=\varphi_7.
\end{cases}
\end{equation}
which is solved by
\begin{equation}\label{eq:dP5-thetas}
\varphi_1=\varphi_2=\varphi_5=\varphi_6=-\varphi_3=-\varphi_4=-\varphi_7=-\varphi_8.
\end{equation}
These constraints leave room for precisely two configurations of central charges within the positive half-plane, namely 
\be\label{eq:dP5-varphi}
	\CC_1[dP_5] : 
	\quad
	\varphi_1 > \varphi_3\,,
	\qquad\qquad
	\CC_2[dP_5] : 
	\quad
	\varphi_1 < \varphi_3\,.
\ee
The corresponding ray diagrams are shown in Figure \ref{Fig:dP5Stability}. 

So far we applied the criteria of Section \ref{Sec:StabRoots}  to  deduce the  stability  conditions associated to flows $T_1$ and $T_2$. 
As a  check, we observe that iterations of  these translation  generators indeed admit a realization that only involves left-mutations 
\be
\begin{split}
	T_1&=\textbf{m}_1:= \mu_3\mu_4\mu_7\mu_8\mu_1\mu_2\mu_5\mu_6\,,
	\\ 
	T_2&=\textbf{m}_2:=\mu_1\mu_2\mu_5\mu_6\mu_3\mu_4\mu_7\mu_8\,.
\end{split}
\ee
Moreover we claim that iterations of these mutations  correspond to tiltings of the positive half-plane for choices of stability data as given  in  (\ref{eq:dP5-varphi}).

\paragraph{BPS spectrum and Wall-crossing invariant.}
A configuration of central charges 
satisfying conditions (\ref{eq:dP5-thetas})-(\ref{eq:dP5-varphi}) 
corresponding to $T_1$ can be taken as follows
\be\label{eq:dP5-central-charges}
\begin{array}{cc}
	\CC_1[dP_5]\supset
	\qquad
	&
	\begin{split}
	&Z_{\gamma_2}=Z_{\gamma_6}=c\cdot Z_{\gamma_1}=c\cdot Z_{\gamma_5}\,,
	\quad
	Z_{\gamma_4}=Z_{\gamma_8}=c\cdot Z_{\gamma_3}=c\cdot Z_{\gamma_7}\,,
	\\
	&
	Z_{\gamma_1}=\overline{Z}_{\gamma_3}\,,
	\quad
	\arg Z_{\gamma_1} > \arg Z_{\gamma_3}\,,
	\end{split}
\end{array}
\ee
with an arbitrary real constant $c>1$, needed to ensure that $Z_{\alpha_i}$ belong to the positive half-plane.
Likewise, a configuration of central charges in chamber $\CC_2[dP_5]$ is relatetd to $\CC_1[dP_5]$ by switching the last condition with $\arg Z_{\gamma_1} < \arg Z_{\gamma_3}$.

Focusing on $\CC_1[dP_5]$, and tilting the positive half-plane of central charges clockwise, induces infinite iterations of the sequence of mutations $\textbf{m}_1$.
Likewise, tilting CCW we find an infinite repetition of the following sequence
\be
	\widetilde{\bf {m}}_1 := \widetilde\mu_6\widetilde\mu_5\widetilde\mu_2 \widetilde\mu_1\widetilde\mu_8\widetilde\mu_7\widetilde\mu_4\widetilde\mu_3\,.
\ee

Together the mutation sequences ${\bf m}_1, \widetilde{\bf m}_1$ generate towers of hypermultiplets
\be
\begin{split}
	\Omega(\gamma_1+n(\gamma_1+\gamma_2+\gamma_3+\gamma_4)) & = 1  \\
	\Omega(\gamma_2+n(\gamma_1+\gamma_2+\gamma_3+\gamma_4)) & = 1  \\
	\Omega(\gamma_3+n(\gamma_1+\gamma_2+\gamma_3+\gamma_4)) & = 1  \\
	\Omega(\gamma_4+n(\gamma_1+\gamma_2+\gamma_3+\gamma_4)) & = 1  \\
	\Omega(\gamma_1+\gamma_2+\gamma_3+n(\gamma_1+\gamma_2+\gamma_3+\gamma_4)) & = 1  \\
	\Omega(\gamma_1+\gamma_2+\gamma_4+n(\gamma_1+\gamma_2+\gamma_3+\gamma_4)) & = 1  \\
	\Omega(\gamma_3+\gamma_4+\gamma_1+n(\gamma_1+\gamma_2+\gamma_3+\gamma_4)) & = 1  \\
	\Omega(\gamma_3+\gamma_4+\gamma_2+n(\gamma_1+\gamma_2+\gamma_3+\gamma_4)) & = 1  \\
\end{split}
\qquad
\begin{split}
	\Omega(\gamma_5+n(\gamma_5+\gamma_6+\gamma_7+\gamma_8)) & = 1  \\
	\Omega(\gamma_6+n(\gamma_5+\gamma_6+\gamma_7+\gamma_8)) & = 1  \\
	\Omega(\gamma_7+n(\gamma_5+\gamma_6+\gamma_7+\gamma_8)) & = 1  \\
	\Omega(\gamma_8+n(\gamma_5+\gamma_6+\gamma_7+\gamma_8)) & = 1  \\
	\Omega(\gamma_5+\gamma_6+\gamma_7+n(\gamma_5+\gamma_6+\gamma_7+\gamma_8)) & = 1  \\
	\Omega(\gamma_5+\gamma_6+\gamma_8+n(\gamma_5+\gamma_6+\gamma_7+\gamma_8)) & = 1  \\
	\Omega(\gamma_7+\gamma_8+\gamma_5+n(\gamma_5+\gamma_6+\gamma_7+\gamma_8)) & = 1  \\
	\Omega(\gamma_7+\gamma_8+\gamma_6+n(\gamma_5+\gamma_6+\gamma_7+\gamma_8)) & = 1  \\
\end{split}
\ee
Since $
	Z_{\gamma_1+\dots+\gamma_4}=Z_{\gamma_5+\dots+\gamma_8}\in \IR^+
$
the BPS rays of these hypermultiplets asymptote to $\IR^+$ form both sides, confirming that (\ref{eq:dP5-central-charges}) defines a collimation  chamber.
It should be noted  that, although many central charges have the same phase, this does not correspond to a wall of marginal stability because all relevant charges are mutually local.

\be
\begin{split}
	\IU(\measuredangle^+,\CC_1) 
	& = 
	\prod^{\nearrow}_{k\geq 0}
	\left[
	\(
		\prod_{\ell=1,2}\Phi\(X_{\gamma_\ell+ k (\gamma_1+\gamma_2+\gamma_3+\gamma_4)}\) 
	\)
	\(
		\prod_{\ell=5,6}\Phi\(X_{\gamma_\ell+ k (\gamma_5+\gamma_6+\gamma_7+\gamma_8)}\) 
	\)
	\right.
	\\
	&
	\times
	\left.
	\(
		\prod_{\ell=3,4}\Phi\(X_{\gamma_1+\gamma_2+\gamma_\ell+ k (\gamma_1+\gamma_2+\gamma_3+\gamma_4)}\) 
	\)
	\(
		\prod_{\ell=7,8}\Phi\(X_{\gamma_5+\gamma_6+\gamma_\ell+ k (\gamma_5+\gamma_6+\gamma_7+\gamma_8)}\) 
	\)
	\right]
	\\ 
	\IU(\measuredangle^-,\CC_1) 
	& = 
	\prod^{\searrow}_{k\geq 0}
	\left[
	\(
		\prod_{\ell=1,2}\Phi\(X_{\gamma_3+\gamma_4+\gamma_\ell+ k (\gamma_1+\gamma_2+\gamma_3+\gamma_4)}\) 
	\)
	\(
		\prod_{\ell=5,6}\Phi\(X_{\gamma_7+\gamma_8+\gamma_\ell+ k (\gamma_5+\gamma_6+\gamma_7+\gamma_8)}\) 
	\)
	\right.
	\\
	&
	\times
	\left.
	\(
		\prod_{\ell=3,4}\Phi\(X_{\gamma_\ell+ k (\gamma_1+\gamma_2+\gamma_3+\gamma_4)}\) 
	\)
	\(
		\prod_{\ell=7,8}\Phi\(X_{\gamma_\ell+ k (\gamma_5+\gamma_6+\gamma_7+\gamma_8)}\) 
	\)
	\right]
\end{split}
\ee

Following the familiar script already used for $\IF_0$ and $dP_3$, to determine $\IU(\IR^+)$ we adopt a formal series expansion as in (\ref{U-real-expansion}) and 
express $\IU$ as a function of the coefficients $c_\gamma$ of $\IU(\IR^+)$ through (\ref{eq:collimation-factorization}). 
We then impose (\ref{eq:symmetry-equation}) using the generator $\pi$ of the $\IZ_4\subset\Pi_Q$ symmetry\footnote{Note that $(\IZ_2)^{\times 4}$ acts trivially on $\CC_1$. Moreover, only a $\IZ_2$ subgroup of $\IZ_4$ acts nontrivially on $\CC_1$, exchanging stability  conditions in Figure \ref{Fig:dP5Stability}.} 
of the quiver of Figure \ref{Fig:QuiverdP5}  
Solving for the coefficients $c_\gamma$ we obtain a series in $X_\gamma$ up to arbitrary order $|\gamma|$, which is compatible with the factorization
\be
	\IU(\IR^+) = \prod_{\gamma} \prod_{m\in \IZ} \Phi((-y)^m X_{\gamma})^{a_m(\gamma)}
\ee
where $\Omega(\gamma,y) = \sum_m (-y)^m a_{m}(\gamma)$  are as follows 
\be
\begin{split}
	\Omega(\gamma_1+\gamma_3 + n\gamma_{D0}) =\Omega(-\gamma_1-\gamma_3 + (n+1)\gamma_{D0})  & =  1\\
	\Omega(\gamma_1+\gamma_4 + n\gamma_{D0}) = \Omega(-\gamma_1-\gamma_4 + (n+1)\gamma_{D0})& =  1\\
	\Omega(\gamma_2+\gamma_3 + n\gamma_{D0}) = \Omega(-\gamma_2-\gamma_3 + (n+1)\gamma_{D0})& =  1\\
	\Omega(\gamma_2+\gamma_4 + n\gamma_{D0}) = \Omega(-\gamma_2-\gamma_4 + (n+1)\gamma_{D0}) & =  1\\
	\Omega(\gamma_5+\gamma_7 + n\gamma_{D0}) = \Omega(-\gamma_5-\gamma_7 + (n+1)\gamma_{D0})  & =  1\\
	\Omega(\gamma_5+\gamma_8 + n\gamma_{D0}) = \Omega(-\gamma_5-\gamma_8 + (n+1)\gamma_{D0})& =  1\\
	\Omega(\gamma_6+\gamma_7 + n\gamma_{D0}) = \Omega(-\gamma_6-\gamma_7 + (n+1)\gamma_{D0}) & =  1\\
	\Omega(\gamma_6+\gamma_8 + n\gamma_{D0}) = \Omega(-\gamma_6-\gamma_8 + (n+1)\gamma_{D0})& = 1\\
	\Omega(\gamma_1+\gamma_2+\gamma_3+\gamma_4 + n\gamma_{D0}) = \Omega(-\gamma_1-\gamma_2-\gamma_3-\gamma_4 + (n+1)\gamma_{D0})& =  y+y^{-1}
\end{split}
\ee
Once again this misses the contributions from pure D0 branes. We refer to the  remarks below (\ref{eq:F0-real-ray-states}) for a discussion.
Based on  the observation that the moduli space  of a D0 brane (or a boundstate  of $n$ D0 branes) is the  Calabi-Yau itself, we expect that once again the Protected Spin Character of $n$ D0 equals the Poincar\'e polynomial of the local del Pezzo~\cite{Mozgovoy:2020has}. Recalling from Section \ref{sec:charge-geom} that $\dim H^2(S,\IZ) = n+1$ we conjecture
\be
	\Omega(n\gamma_{D0};y) = y^3 + 6y+y^{-1}\,.
\ee

In conclusion, the spectrum we obtain for local $dP_5$ in chamber  $\CC_1$ is 
\begin{equation}\label{eq:SpectrumdP5}
\begin{array}{|c|c|}
	\hline
	\gamma & \Omega(\gamma;y) \\
	\hline\hline
	\gamma_r + k (\gamma_1+\gamma_2+\gamma_3+\gamma_4) & 1\\
	-\gamma_r + (k+1) (\gamma_1+\gamma_2+\gamma_3+\gamma_4) & 1\\
	\gamma_s + k (\gamma_5+\gamma_6+\gamma_7+\gamma_8) & 1\\
	-\gamma_s + (k+1) (\gamma_5+\gamma_6+\gamma_7+\gamma_8) & 1\\
	\hline
	\gamma_a+\gamma_b + k \gamma_{D0} & 1\\
	-\gamma_a-\gamma_b + (k+1) \gamma_{D0} & 1\\
	\gamma_1+\gamma_2+\gamma_3+\gamma_4 + k \gamma_{D0} & y+y^{-1}\\
	-\gamma_1-\gamma_2-\gamma_3-\gamma_4 + (k+1) \gamma_{D0} & y+y^{-1}\\
	(k+1)\gamma_{D0} & y^3 + 6y+y^{-1}\\
	\hline
\end{array}
\end{equation}
with $k\geq  0$ and 
\be\label{eq:dP5-charges-domains}
\begin{split}
	& r\in \{1,2,3,4\},\quad s\in \{5,6,7,8\}  \\
	& (a,b)\in \{(1,3), (1,4), (2,3), (2,4), (5,7), (5,8), (6,7), (6,8)\} 
	\\
\end{split}
\ee
The spectrum also  includes the respective antiparticles obtained by sending $\gamma\to-\gamma$.

\begin{remark}
The wall-crossing constraint that we solved is the following identity 
\be
\begin{split}
	& 
	\prod^{\nearrow}_{k\geq 0}
	\left[
	\(
		\prod_{\ell=1,2}\Phi\(X_{\gamma_\ell+ k (\gamma_1+\gamma_2+\gamma_3+\gamma_4)}\) 
	\)
	\(
		\prod_{\ell=5,6}\Phi\(X_{\gamma_\ell+ k (\gamma_5+\gamma_6+\gamma_7+\gamma_8)}\) 
	\)
	\right.
	\\
	&
	\times
	\left.
	\(
		\prod_{\ell=3,4}\Phi\(X_{\gamma_1+\gamma_2+\gamma_\ell+ k (\gamma_1+\gamma_2+\gamma_3+\gamma_4)}\) 
	\)
	\(
		\prod_{\ell=7,8}\Phi\(X_{\gamma_5+\gamma_6+\gamma_\ell+ k (\gamma_5+\gamma_6+\gamma_7+\gamma_8)}\) 
	\)
	\right]
	\\ 
	&
	\prod_{k\geq 0}
	\Bigg[
	\(
		\prod_{s=\pm1} \Phi\((-y)^s X_{\gamma_1+\gamma_2+\gamma_3+\gamma_4+ k \gamma_{D0}}\)^{-1}  
		\Phi\((-y)^s X_{-\gamma_1-\gamma_2-\gamma_3-\gamma_4+ (k+1) \gamma_{D0}}\)^{-1} 
	\)
	\\
	&\qquad\times
	\(
		\prod_{(a,b)} 
		\Phi\(X_{\gamma_a+\gamma_b+ k \gamma_{D0}}\)  
		\Phi\(X_{-\gamma_a-\gamma_b+ (k+1) \gamma_{D0}}\) 
	\) 
	\Bigg]
	\\
	& 
	\times 
	\prod^{\searrow}_{k\geq 0}
	\left[
	\(
		\prod_{\ell=1,2}\Phi\(X_{\gamma_3+\gamma_4+\gamma_\ell+ k (\gamma_1+\gamma_2+\gamma_3+\gamma_4)}\) 
	\)
	\(
		\prod_{\ell=5,6}\Phi\(X_{\gamma_7+\gamma_8+\gamma_\ell+ k (\gamma_5+\gamma_6+\gamma_7+\gamma_8)}\) 
	\)
	\right.
	\\
	&
	\times
	\left.
	\(
		\prod_{\ell=3,4}\Phi\(X_{\gamma_\ell+ k (\gamma_1+\gamma_2+\gamma_3+\gamma_4)}\) 
	\)
	\(
		\prod_{\ell=7,8}\Phi\(X_{\gamma_\ell+ k (\gamma_5+\gamma_6+\gamma_7+\gamma_8)}\) 
	\)
	\right]
	\\
	& =  
	\(\{\gamma_1,\gamma_2,\gamma_3,\gamma_4,\gamma_5,\gamma_6,\gamma_7,\gamma_8\}\to\{\gamma_3,\gamma_4,\gamma_5,\gamma_6,\gamma_7,\gamma_8,\gamma_1,\gamma_2\}\)
\end{split}\label{eq:WCIDP5}
\ee
where labels $(a,b)$ run over the same combinations as in (\ref{eq:dP5-charges-domains}).
As we did previously, we factored out the contribution of  pure-flavor states (pure D0 branes), which  do not participate in wall-crossing. We have tested this as an identity in the numerical ($y\to -1$) setting, by acting  on $X_{\gamma_i}$ and expanding as formal series in $X_\gamma$ with $|\gamma|\leq 10$.
\end{remark}
\begin{figure}[h]
\begin{center}
\begin{subfigure}{.35\textwidth}
\centering
\includegraphics[width=\textwidth]{figures/dP5Sub.jpg}
\caption{Four-dimensional subquivers for the flow $T_1$ of $dP_5$.}
\label{Fig:dP51}
\end{subfigure}\hspace*{.15\textwidth}
\begin{subfigure}{.35\textwidth}
\centering
\includegraphics[width=\textwidth]{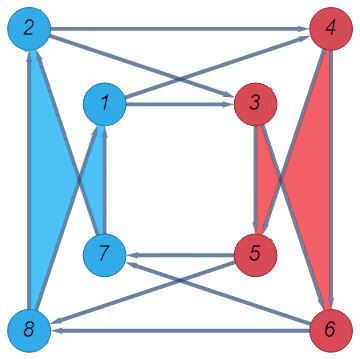}
\caption{Four-dimensional subquivers for the flow $T_2$ of $dP_5$.}
\label{Fig:dP52}
\end{subfigure}
\caption{}
\label{fig:F0-subquivers}
\end{center}
\end{figure}

\bibliography{biblio}{}
\bibliographystyle{JHEP}

\end{document}